\documentclass[ twocolumn]{aastex631}
\usepackage{amsmath}
\usepackage{CJK}
\usepackage{xcolor}
\usepackage{graphicx}

\usepackage{booktabs} % For nicer lines in tables.

\usepackage{svg} % For nicer lines in tables.
\usepackage{float}

\usepackage{bm}
\usepackage{soul}

\defcitealias{goldreich_toward_1995}{GS95}
\defcitealias{boldyrev_2006}{B06}
\defcitealias{chandran_intermittency_2015}{CSM15}
\defcitealias{Mallet_2017}{MS17}
\defcitealias{iroshnikov_turbulence_1963}{IK}
\defcitealias{kolmogorov_local_1941}{K41}

\definecolor{airforceblue}{rgb}{0.36, 0.54, 0.66}
\definecolor{bluegray}{rgb}{0.4, 0.6, 0.8}

\usepackage{hyperref, array}

\begin{document}
% \begin{CJK*}{UTF8}{gbsn}

\shorttitle{3D anisotropy of Imbalanced Turbulence}
\shortauthors{Sioulas et al.}

\correspondingauthor{Nikos Sioulas}
\email{nsioulas@g.ucla.edu}

\title{Higher-Order Analysis of Three-Dimensional Anisotropy in Imbalanced Alfv\'enic Turbulence}%observed by Parker Solar Probe

\author[0000-0002-1128-9685]{Nikos Sioulas}
\affiliation{Department of Earth, Planetary, and Space Sciences, University of California, Los Angeles, CA, USA }

\author[0009-0000-4345-2486]{Themistocles Zikopoulos}
\affiliation{Department of Physics, Aristotle University of Thessaloniki, GR-52124 Thessaloniki, Greece}

\author[0000-0002-2582-7085]{Chen Shi}
\affiliation{Department of Earth, Planetary, and Space Sciences, University of California, Los Angeles, CA, USA}

\author[0000-0002-2381-3106]{Marco Velli}
\affiliation{Department of Earth, Planetary, and Space Sciences, University of California, Los Angeles, CA, USA}

\author[0000-0002-4625-3332]{Trevor Bowen}
\affiliation{Space Sciences Laboratory, University of California, Berkeley, CA 94720-7450, USA}

\author[0000-0003-4177-3328]{Alfred Mallet}
\affiliation{Space Sciences Laboratory, University of California, Berkeley, CA 94720-7450, USA}

\author[0000-0002-5981-7758]{Luca Sorriso-Valvo}
\affiliation{Institute for Plasma Science and Technology (ISTP), CNR, Bari, Italy}
\altaffiliation{Space and Plasma Physics, School of Electrical Engineering and Computer Science, KTH Royal Institute of Technology, Stockholm, Sweden}

\author[0000-0003-4380-4837]{Andrea Verdini}
\affiliation{Universit\'a di Firenze, Dipartimento di Fisica e Astronomia, Firenze, Italy}

\author[0000-0003-4177-3328]{B. D. G. Chandran}
\affiliation{Space Science Center and Department of Physics, University of New Hampshire, Durham, NH 03824, USA}

\author[0000-0002-7365-0472]{Mihailo M. Martinovi\'c}
\affiliation{Lunar and Planetary Laboratory, University of Arizona, Tucson, AZ 85721, USA}

\author[0000-0003-0562-6574]{S. S. Cerri}
\affiliation{Université Côte d’Azur, Observatoire de la Côte d’Azur, CNRS, Laboratoire Lagrange, Bd de l’Observatoire, CS 34229, 06304 Nice5cedex 4, France}

\author[0000-0001-7222-3869]{Nooshin Davis}
\affiliation{Space Science Center and Department of Physics, University of New Hampshire, Durham, NH 03824, USA}

\author[0009-0007-8306-2635]{Corina Dunn}
\affiliation{Space Sciences Laboratory, University of California, Berkeley, CA 94720-7450, USA}

\begin{abstract}

In-situ observations of imbalanced solar wind turbulence are employed to evaluate the core assumptions and higher-order scaling predictions of MHD turbulence models grounded in the principles of ``Critical Balance'' (CB) and ``Scale-Dependent Dynamic Alignment'' (SDDA).

Our results indicate that, at the energy injection scales, both outgoing and ingoing Alfv\'enic fluctuations undergo a weak energy cascade, $\chi_{\lambda}^{\pm} < 1$, where $\chi_{\lambda}^{\pm} \equiv \tau_{A}^{\pm}/\tau_{nl}^{\pm}$, the ratio of linear to non-linear timescales. Simultaneously, a tightening of SDDA is observed across this range. While the outgoing waves remain in a weak cascade state throughout the inertial range, $\chi_{\lambda}^{+} \approx 0.2$, the ingoing waves transition to a strong cascade,  $\chi_{\lambda}^{-} > 1$, at $\lambda \approx 3 \times 10^{4} d_i$. This transition, however, is accompanied by spectral scalings that diverge from the expected canonical behavior marking the shift from weak to strong turbulence - a discrepancy we attribute to the effects of ``anomalous coherence''. The domain canonically identified as the inertial range consists of two distinct sub-inertial segments. At $\lambda \gtrsim 100 d_i$, the ``average'' eddy assumes a field-aligned tube topology, with SDDA signatures being weak and largely restricted to the highest amplitude fluctuations. The scaling exponents $\zeta_{n}$ of the $n$-th order conditional structure functions, perpendicular to both the local mean field and the fluctuation direction, conform to the analytical models of \cite{Chandran_2015} and \cite{Mallet_2017}, indicating ``multifractal'' statistics and strong intermittency; the scaling in the parallel and displacement (i.e., fluctuation direction) components is more concave than theoretically predicted.  We argue that the statistics of this range could be contaminated by expansion effects. Below $\lambda \approx 100 d_i$, eddies display increasing anisotropy, evolving into structures resembling thin current sheets. Concurrently, $\zeta_{n}$ scales linearly with order, signaling a transition towards ``monofractal'' statistics. At $\lambda \approx 8 d_i$, the increase in aspect ratio ceases, and the eddies transition to a quasi-isotropic state. This shift might be a signature of the tearing instability, potentially leading to reconnection of the thin current sheets, or it could result from turbulent energy being channeled into an ion-cyclotron wave spectrum, consistent with the ``helicity barrier'' effect.

Our analysis employs 5-point structure functions, which are shown to be more effective than the traditional 2-point approach in accurately capturing the steep scaling behaviors observed at smaller spatial scales.

\end{abstract}


\section{Introduction}\label{sec:intro}

Propelled by the internal dynamics of the Sun, the solar wind expands spherically within the heliosphere, transporting a wide spectrum of magnetic field and plasma fluctuations \citep{bruno_solar_2013}. Given that fluctuations are predominantly weakly compressive \citep{Tu95}, their dynamics are often analyzed within the framework of incompressible magnetohydrodynamic (MHD) theory \citep{Biskamp03, Matthaeus11}. The \cite{elsasser_1950} form of the ideal incompressible MHD equations is given by 
\begin{equation}\label{eq:intro1}
    \frac{\partial \delta \boldsymbol{z}^{\pm}}{\partial t} \mp (\boldsymbol{V_{A}} \cdot \nabla)  \delta \boldsymbol{z}^{\pm} = -(  \delta \boldsymbol{z}^{\mp} \cdot \nabla)  \delta \boldsymbol{z}^{\pm} - \frac{\nabla p}{\rho_0},
\end{equation}
where the Els\"asser variables, $\delta \boldsymbol{z}^{\pm} = \delta \boldsymbol{v} \pm \delta \boldsymbol{b}/ \sqrt{4 \pi \rho}$, represent eigenfunctions of \cite{Alfven_1942} waves propagating (anti)parallel to the background magnetic field ($\boldsymbol{B}_{0}$) at the Alfv\'en speed, $\boldsymbol{V}_{a} = \boldsymbol{B}_{0}/ \sqrt{4 \pi \rho}$. The total energy $E_{t} = E^{+} + E^{-}$, and cross-helicity $H_{c} = E^{+} - E^{-}$, expressed in terms of the energy associated with fluctuations in $\boldsymbol{z}^{\pm}$,  $E^{\pm} = \langle |\delta \boldsymbol{z}^{\pm}|^{2}\rangle/4 $  are ideal (i.e., with zero viscosity and resistivity) invariants of the incompressible MHD equations. Their ratio defines the normalized cross helicity $\sigma_c  = H_c/E_t$. When the energy fluxes, denoted as $\epsilon^{\pm}$, in wave packets traveling in opposite directions differ, indicated by $\epsilon^{+}/\epsilon^{-} \neq 1$, turbulence is termed as $\textit{imbalanced}$, $\sigma_c \neq 0$.  
The nonlinearities, as highlighted by the first term on the right-hand side of Equation \ref{eq:intro1}, arise from collisions between oscillation modes with opposite signs of cross-helicity (i.e. between  $\delta \boldsymbol{z}^{+}$ and $\delta \boldsymbol{z}^{-}$). These inertial, or energy-conserving, interactions lead to the fragmentation of the wavepackets and drive energies
of $E^{+}$ and $E^{-}$ towards smaller perpendicular scales, at which point energy is converted into heat through dissipation \citep{iroshnikov_turbulence_1963, kraichnan_inertial-range_1965}, henceforth referred to as \citetalias{iroshnikov_turbulence_1963}. The turbulent cascade is strongly anisotropic and results in the formation of structures characterized by $\ell_{||} \gg \lambda$, with $\ell_{||} \sim 1/ k_{||}$ and $\lambda \sim 1/ k_{\perp}$ representing the correlation lengths along and perpendicular to the background magnetic field\footnote{In assessing the dominant scale of the background magnetic field, two methodologies are predominantly employed, utilizing the global and local frames \citep{Cho_Vishniac_2000, Maron_2001}.  For a detailed exploration of the consequences associated with defining the magnetic field either globally, thereby $\boldsymbol{B}_{0}$, or locally, denoted as $\boldsymbol{B}_{\ell}$, interested readers are encouraged to refer to \citep{Chen-anisotropic_2011, Matthaeus_2012_local_global, Gerick_2017}.}, respectively \citep{Montgomery_1981, shebalin_matthaeus_montgomery_1983, Grappin_1986}. The effectiveness of the collisions hinges on two critical timescales: the time it takes for a turbulent perturbation of size $\lambda$ to break up nonlinearly $\tau_{nl}^{\pm} \sim \lambda/ \delta z^{\mp}$  and the linear propagation (or collision) time $\tau_{A}^{\pm} = \ell_{||}^{\pm}/ V_a$. The ratio of these timescales defines the nonlinearity parameter, $\chi^{\pm} \equiv \tau_{A}^{\pm}/\tau_{nl}^{\pm} \sim (\ell_{\lambda}^{\pm}/ \lambda)(\delta \boldsymbol{z}^{\pm}_{\lambda} / V_{a})$.\footnote{We assumed
here, implicitly, that the cascade is local in $k_{\perp}$. However, it's important to note that this condition may not hold in cases of strongly imbalanced turbulence \citep{Schekochihin_2022}. This point is further discussed in Section \ref{sec:Disussion}.}

 In the limit of weak, turbulence,   $\chi \ll 1$, energy transfer to smaller scales occurs fractionally upon each collision, with weak nonlinear effects accumulating over timescales significantly longer than the wave period. Under the assumption that collisions accumulate in a manner akin to a standard random walk, approximately $N \sim 1/\chi^{2}$ collisions are necessary to fully cascade the energy at a given scale. 
Dimensional analysis \citep{1996_Ng_Bhattacharjee} and the formal application of weak perturbative theory \citep{Galtier_2000} both predict an inertial range spectrum scaling as $E \propto k_{\perp}^{-2}$.
\par An initial state of globally weak turbulence is often unstable, and the 
intrinsic anisotropy of the energy cascade inevitably develops smaller perpendicular scales that are strongly turbulent with nonlinear effects present at the leading order. The case of strong turbulence was addressed by \cite{higdon_anisotropic} and \cite{goldreich_toward_1995}, hereafter \citetalias{goldreich_toward_1995}. 
\citetalias{goldreich_toward_1995} expanded upon the idea that fluctuations in any two planes perpendicular to the mean field can remain correlated only if an Alfv\'en wave can propagate between them in less time than their perpendicular decorrelation time \citep[see also, ][]{Maron_2001, schekochihin_astrophysical_2009}. This leads to the implication that the dynamics of the inertial range in incompressible MHD turbulence are governed by wavevector modes in a ``critical balance'' (CB) state, i.e., characterized by a near-equal balance between the two dynamically important timescales, essentially achieving $\chi \sim 1$. Under the assumption of negligible cross helicity, which suggests identical statistical properties for counterpropagating wavepackets, this implies that the parallel and perpendicular wavevectors follow the scaling law $k_{||} \propto k_{\perp}^{2/3}$. Consequently, the one-dimensional power spectra for total energy scale as $E(k_{\perp}) \propto k_{\perp}^{-5/3}$ and $E(k_{||}) \propto k_{||}^{-2}$, across and along the local magnetic field, $\boldsymbol{B}_{\ell}$, respectively.

\par The \citetalias{goldreich_toward_1995} model is directly applicable to the Reduced MHD approximation (RMHD), where the background magnetic field $\boldsymbol{B}_{0}$ is significantly stronger than the fluctuating amplitudes \citep{Kadomtsev_Pogutse, Strauss_1976, Oughton_2017ApJ}. The latter are restricted to a plane orthogonal to $\boldsymbol{B}_0$. Despite $\delta \boldsymbol{b}_{\perp} \ll \boldsymbol{B}_0$, nonlinear effects are retained at the leading order. This is achieved by excluding all high-frequency fluctuations $\tau_{a} \leq \tau_{nl}$. Consequently, fluctuations within the RMHD approximation inherently satisfy the condition $\chi \ge 1$ \citep{Oughton_2020_cb}. In this case, nonlinear interactions along the mean field may be completely neglected, allowing the nonlinear evolution to adhere to the 2D incompressible MHD equations in planes orthogonal to $\boldsymbol{B}_{0}$\footnote{Note, however, that the RMHD approximation encompasses essential elements of the physics of three-dimensional incompressible MHD \citep{Dmitruk_2005, Oughton_2017_RMHD}}.

\par While the core principle of the \citetalias{goldreich_toward_1995}  model, namely CB, was shown to be consistent with numerical simulations of homogeneous, (in)compressible MHD turbulence, the inertial range scaling perpendicular to $\boldsymbol{B}_{\ell}$ was observed to be closer to -3/2 \citep{Maron_2001, 2003PhRvE..67f6302M, Muller_grappin_2005}\footnote{Recent higher resolution simulations further support this finding \citep{Perez_2012, 2015_verdini, 2016_mallet, Dong_22_largest_mhd_turb, 2023_Chen_compres}}. Additionally, numerical simulations revealed a tendency for magnetic and velocity fluctuations in the field-perpendicular plane to align with each other within a small, scale-dependent angle \citep{Beresnyak_2006}.  

\par To reconcile the noted discrepancy, \cite{Boldyrev_2005, boldyrev_2006}, henceforth \citetalias{boldyrev_2006}, proposed a phenomenological model linking the emergence of local imbalance \citep{dobrowolny_fully_1980, Matthaeus_2008_patches_imbalance} with the scale-dependent dynamic alignment (SDDA) in the polarizations of $\delta\boldsymbol{b}_{\perp}$ and $\delta\boldsymbol{v}_{\perp}$, towards smaller scales, $\theta_{\perp}^{ub} \sim \delta b/v_A \propto \lambda^{1/4}$. Drawing on geometrical considerations, \citetalias{boldyrev_2006} suggests that the observed increase in alignment at smaller scales is linked to both a depletion of nonlinearities and simultaneous development of local anisotropy in the plane perpendicular to $\boldsymbol{B}_{\ell}$. In this framework turbulent eddies are identified as 3D-anisotropic structures, characterized by $\ell_{||} \gg \xi \gg \lambda$, where, $\lambda/\xi \sim \sin \theta_{\perp}^{ub}$. Here, $\xi$ represents the coherence length in the direction of $\delta \boldsymbol{b}$. \citetalias{boldyrev_2006} predicts three distinct scaling laws in the inertial range: $E(k_{\xi}) \propto k_{\xi}^{-5/3}$, $E(k_{\lambda}) \propto k_{\lambda}^{-3/2}$, and $E(k_{\ell_{||}}) \propto k_{\ell_{||}}^{-2}$. The \citetalias{boldyrev_2006} model has been subject to criticism by \cite{Beresnyak_2011}, on the basis that it conflicts with the rescaling symmetry intrinsic to the RMHD equations.

\par

The concept of SDDA received further refinement in the work of \citet{chandran_intermittency_2015}, hereinafter referred to as \citetalias{chandran_intermittency_2015}. In their interpretation, alignment emerges as an intermittency effect, resulting from the mutual shearing of Els\"asser fields during the imbalanced collisions ($\delta z^{\pm} \gg \delta z^{\mp}$) of counterpropagating wave packets. Furthermore, \citet{Mallet_2017}, subsequently referred to as \citetalias{Mallet_2017}, formulated a statistical model for three-dimensional 3D anisotropic, intermittent Alfv\'enic turbulence. Both models integrate the principles of CB and SDDA, preserving the scale invariance characteristic of RMHD.
With the incorporation of SDDA, the nonlinear timescale is defined as $\tau_{nl}^{\pm} \sim \lambda / (\delta z^{\mp} \sin\theta^{z})$, wherein $\theta^{z}$ denotes the angle between the fluctuations of the Els\"asser variables in the plane perpendicular to the magnetic field, $\delta z^{\pm}_{\perp}$. \citetalias{Mallet_2017} extends predictions for 3D anisotropic higher-order scaling that are consistent with the second-order moment scaling proposed by \citetalias{boldyrev_2006}.

\par The aforementioned theoretical models focus on the dynamics of small amplitude (toroidal) Alfv\'en modes, neglecting potential couplings with compressive fluctuations \citep[e.g., ][]{Cho_Lazarian_2003, Chandran_2005, chandran_2018}, within the context of globally $\textit{balanced}$ turbulence. In contrast, solar wind turbulence typically displays a predominance in the flux of outwardly propagating fluctuations over inwardly directed ones 
\citep{Roberts_1987, 2021A&A...654A.111D}. In addition, the predominant fluctuations in the solar wind are consistent with large amplitude ($\delta \boldsymbol{b} \sim \boldsymbol{B}_0$) Alfv\'en waves. These waves are not purely transverse to $\boldsymbol{B}_0$ but exhibit spherical polarization, meaning the magnetic field vector traces a sphere of constant radius $|\boldsymbol{B}_0 + \delta \boldsymbol{b}| = const.$ \citep{barnes_large-amplitude_1974, Bruno_2001,
Matteini_2014}. Furthermore, a non-negligible fraction of compressive fluctuations is observed  \citep{Yao_2011, howes_slow-mode_2012, Klein_2012}. Finally, the dynamics of the solar wind are influenced not just by the mean field direction, but also by the radial axis along which the solar wind expands \citep{1973_Volk}. This aspect has been illuminated by numerical simulations using the Expanding Box Model (EBM) \citep{Grappin_velli_mengeney_1993, 1996_grappin_velli}, which demonstrate that expansion preferentially reduces the radial component of the magnetic field across all scales, confining fluctuations to a plane orthogonal to the radial direction \citep{Dong_2014}. Given these considerations, the extent to which phenomenological models of homogeneous,  Alfv\'enic turbulence can accurately capture the unique characteristics of the solar wind remains an active area of debate \citep{bowen2021nonlinear}.

\par In this investigation, we endeavor to evaluate the consistency of the homogeneous models of balanced MHD turbulence proposed by \citetalias{chandran_intermittency_2015} and \citetalias{Mallet_2017} against  in-situ observations sampled during the first perihelion (E1) of the Parker Solar Probe mission, \citep[PSP,][]{fox_solar_2016}. Our primary objective is to conduct a rigorous comparison between the predicted scalings of higher-order moments in these models and the corresponding empirical observations, with a specific focus on determining the presence and measurable impact of model-specific elements, such as SDDA and CB, on the observed characteristics. Through this analysis, our aim is to not only provide deeper insights into the higher-order statistics of magnetic turbulence but also to establish a robust benchmark for the testing and refinement of theoretical models addressing imbalanced MHD turbulence in the inhomogeneous solar wind.  

\par The structure of the remainder of the paper is as follows: Section \ref{sec:background} provides theoretical background on intermittency and summarizes the core principles of the \citetalias{chandran_intermittency_2015} and \citetalias{Mallet_2017} models. Section \ref{sec:data analysis } elaborates on the methodologies utilized in this analysis. Details regarding data selection and processing are outlined in Section \ref{sec:Data}. The study's findings are presented in Section \ref{sec:Statistical Results}. A comprehensive comparison with prior theoretical, observational, and numerical studies, which contextualizes our results, is provided in Section \ref{sec:Disussion}. The paper concludes with a summary of the main findings in Section \ref{sec:Conclusions}.

\section{Theoretical background: Intermittency}\label{sec:background}

\cite{kolmogorov_local_1941} (hereafter \citetalias{kolmogorov_local_1941}), postulated that the cascade process is intricate enough for eddies to lose all memory of their past and that their properties after each cascade step can be explained by random distributions. An inertial range arises where eddies are too large for viscosity to play a significant role and too small to retain any influence from large-scale inhomogeneities. A universal probability density function (PDF), $P(\delta u, \ell)$, emerges over this range. The velocity difference between non-proximate points can be expressed as the sum of velocity differences over subintervals, justifying a plausible Gaussian distribution assumption based on the central limit theorem. The validity of this assumption hinges on two conditions: mutual independence among summands and comparable finite variances in the subinterval probability distributions \citep{feller1}. Considering a global energy transfer  rate that is independent of scale, $\epsilon$, and utilizing the statistical moments of the PDFs called structure functions ($SF^{n}$):

\begin{equation}
SF_{i}^{n}(\ell)=\int_{-\infty}^{\infty}{(\delta{u}_{i})^nP(\delta{u}_{i}, \ell)d(\delta{u}_{i})},
\label{eq:1}
\end{equation}

where, $\delta{u}_{i}$ denotes longitudinal velocity increments, $\delta u_{i} = V_{i}(\boldsymbol{r} + \boldsymbol{\ell}) - V_{i}(\boldsymbol{r})$, Kolmogorov's similarity hypothesis yields:

\begin{equation} \label{eq:2}
SF^{n}_{i}(\ell) \sim \epsilon^{n/3} \ell^{\zeta^{K41}(n)},
\end{equation}

where $\zeta^{K41}(n) = n/3$, implying global scale invariance (self-similarity) of the fluctuations.

In the case of MHD turbulence, \citetalias{iroshnikov_turbulence_1963} introduced a phenomenological model based on the coupling of small-scale velocity and magnetic fluctuations by the large-scale background magnetic field, $\boldsymbol{B}_{0}$. Compared to \citetalias{kolmogorov_local_1941}, the spectral transfer is reduced by a factor of $\tau_{A}/ \tau_{nl}$ owing to the limited strength of the interaction exhibited by Alfv\'en waves. Also in this case the scaling exponents of $SF^{n}$ are expected to scale linearly with order $n$, i.e., $\zeta^{IK}(n) = n/4$.

However, it was soon realized experimentally that PDFs of fluctuations in hydrodynamic (HD) and MHD turbulence tend to display increasingly non-Gaussian behavior at progressively smaller scales \citep{Batchelor_intermittency, 1991JGR....96.5847B, Sorriso-Valvo99}. Additionally, it has been noted that the spatial inhomogeneity of energy dissipation is expected to alter the scaling exponents of field increments with respect to length scales $\ell$ \citep{1962_Oboukhov, kolmogorov_refinement_1962}. Specifically, if the energy transfer rate from larger to smaller eddies statistically varies with $\ell$, then the dissipation rate $\epsilon$ in Equation \ref{eq:2} should be substituted with $\epsilon_{\ell}$ \citep{1959_Landau}:

\begin{equation}
SF^{n}_{i}(\ell) \sim \langle \epsilon_{\ell}^{n/3}\rangle \ell^{n/3}.
\label{eq:3}
\end{equation}

 Expressing $\epsilon_{\ell}^{n/3}$ through a scaling relation with $\ell$, we find $\langle \epsilon_{\ell}^{n/3}\rangle \sim \ell^{\tau_{n}/3}$, leading to

\begin{equation}
    SF^{n}_{i}(\ell) \sim  \ell^{\zeta_{n}},
    \label{eq:4}
\end{equation} 
where, $\zeta_{n}= n/3 + \tau_{n}/3$ is generally a nonlinear function of n. When analyzing the scaling exponents $\zeta_{n}$ of the structure functions, deviations from linear scaling, indicate a violation of global scale invariance, which implies a process characterized by ``multifractal'' statistics and intermittency, i.e., the concentration of the energy into smaller volumes of space at smaller scales \citep{frisch_turbulence_1995}. Several authors have put forward theoretical arguments to account for the intermittency effect \citep{10.2307/2682923, 1978JFM....87..719F, She_Leveque,  Grauer1994ScalingOH, PhysRevE.52.636,  1995JGR...100.3395R, 1997_horbury_inter, PhysRevLett.84.475, 2002_Boldyrev}. \par

Chandran, Schekochihin, and Mallet \cite{chandran_intermittency_2015} (henceforth \citetalias{chandran_intermittency_2015}) introduced a model founded on collisions of Alfv\'enic wavepackets making a series of plausible assumptions regarding the dynamics and statistics of RMHD turbulence. In formulating an analytical model, they proposed two primary types of non-linear interactions.  The first involves occasional ``balanced'' interactions, $\delta \boldsymbol{z}^{-} \approx \delta \boldsymbol{z}^{+}$, where the amplitude of the wavepackets is reduced by a factor of $0 \leq \beta \leq 1$, yet their length scale remains unchanged. Consequently, the amplitude of a fluctuation at scale $\lambda$ resulting from such balanced collisions can be expressed as follows:

\begin{equation}
\delta \boldsymbol{z}^{\pm}_{\lambda} = \delta \bar{z} \beta^{q}.
\label{eq:5}
\end{equation}

Here, $\delta \bar{z}$ is the initial fluctuation amplitude at the injection scale $L$, and $q$ is the number of balanced collisions expected as the fluctuation evolves from scale $L$ to $\lambda$. The value of $q$ is presumed to follow a Poisson distribution:

\begin{equation}
P(q) = \frac{e^{- \mu} \mu^{q}}{q !},
\label{eq:6}
\end{equation}

where $\mu$ is the scale-dependent mean value of $q$. The ``typical” fluctuation amplitude that best characterizes the bulk of the volume is

\begin{equation}
\delta \boldsymbol{z}^{\ast}_{\lambda}  = \delta \bar{z} \beta^{\mu}.
\label{eq:6a}
\end{equation}

On the contrary, for ``imbalanced'', $\delta \boldsymbol{z}^{\pm} \gg \delta \boldsymbol{z}^{\mp}$, collisions, the amplitudes of the fluctuations remain constant while the sub-dominant field is sheared into alignment and its perpendicular scale $\lambda$ reduces. Assuming that the most intense coherent structures in MHD turbulence --specifically, those with $q=0$-- are 3D anisotropic current sheets with a volume filling factor $f_{cs} \propto \lambda$, this leads to a straightforward relation for the scaling exponents:

\begin{equation}
\zeta^{CSM15}(n)=1-\beta^n.
\label{eq:7}
\end{equation}

By focusing solely on the scenario of zero cross helicity, \citetalias{chandran_intermittency_2015} deduce a value for $\beta \approx 0.691$. This implies that $\zeta_{n} \rightarrow 1$ as $n \rightarrow \infty$.

Furthermore, the model offers predictions for the alignment angles between the perpendicular components of velocity-magnetic and/or Els\"asser field fluctuations defined in various ways. Conventionally, the alignment is estimated as 

\begin{equation}\label{eq:9}
sin (\theta^{ub}_{\perp}) =\left\langle \frac{|\delta \boldsymbol{b}_{\perp} \times \delta\boldsymbol{v}_{\perp}| }{ |\delta \boldsymbol{b}_{\perp}| |\delta \boldsymbol{v}_{\perp}| }\right\rangle.
\end{equation}

 A similar expression can be formulated for the angle between Els\"asser field fluctuations, $\delta \boldsymbol{z}^{\pm}_{\perp}$, which relates to the residual of magnetic over kinetic energy, usually studied through the lens of the normalized residual energy :

 \begin{equation}\label{eq:res_energy}
    \sigma_r = \frac{\langle \delta \boldsymbol{v}_{\perp}  ^{2}\rangle -\langle \delta \boldsymbol{b}_{\perp}  ^{2}\rangle}{ \langle \delta \boldsymbol{v}_{\perp}  ^{2}\rangle  + \langle \delta \boldsymbol{b}_{\perp}  ^{2}\rangle}.
\end{equation}

Similarly, the Els\"asser imbalance is assessed by comparing the relative energy in inwardly and outwardly propagating Alfv\'en waves \citep{Velli_91_waves, Velli_93}:

 \begin{equation}\label{eq:cross_helicity}
   \sigma_c = \frac{2 \langle \delta \boldsymbol{v}_{\perp} \cdot \delta \boldsymbol{b}_{\perp} \rangle}{\langle \delta \boldsymbol{v}_{\perp}^2 \rangle + \langle \delta \boldsymbol{b}_{\perp}^2 \rangle}
\end{equation}

   % \sigma_c =  \frac{\langle \delta \boldsymbol{z}^{+}_{\perp}^{2} - \delta \boldsymbol{z}^{-}_{\perp}  ^{2}\rangle}{\langle  \delta \boldsymbol{z}^{+}_{\perp}  ^{2} + \delta \boldsymbol{z}^{-}_{\perp}  ^{2}\rangle}

Observations from numerical simulations showed little to no scaling of alignment angles as defined in Equation \ref{eq:9}, \citep{Beresnyak_2009B}. However, an alternate definition of these angles, achieved by separately averaging the numerator and denominator --referred to as polarization intermittency-- was proposed by \citep{Beresnyak_2006}:

\begin{equation}\label{eq:10}
sin (\Tilde{\theta}^{ub}_{\perp}) = \frac{\langle |\delta \boldsymbol{b}_{\perp} \times \delta\boldsymbol{v}_{\perp}| \rangle}{ \langle |\delta \boldsymbol{b}_{\perp}| |\delta \boldsymbol{v}_{\perp}| \rangle}.
\end{equation}

In a similar manner, we can define the angle between the two Els\"asser fields as $\Tilde{\theta}^{z}_{\perp}$. The model by \citetalias{chandran_intermittency_2015} predicts  $\Tilde{\theta}^{ub}_{\perp}\propto \lambda^{0.21}$ and $\Tilde{\theta}^{z}_{\perp}\propto \lambda^{0.10}$.

\par \cite{Mallet_2017} (hereafter, \citetalias{Mallet_2017}) formulated a statistical model of RMHD turbulence grounded in three principal concepts: critical balance, dynamic alignment, and intermittency. To substantiate their model, they put forth four conjectures based on physical reasoning: (a) the fluctuation amplitudes adhere to an anisotropic log-Poisson distribution \citep{chandran_intermittency_2015, 2016Zhdankin}; (b) the structures at small scales are 3D anisotropic with a sheet-like morphology \citep{boldyrev_2006, Howes2015}; (c) the scale-independence of the critical balance parameter, inclusive of dynamic alignment, within the inertial range \citep{2015Mallet}; (d) a consistent energy flux across parallel scales in the inertial range \citep{2015Beresnyak_paral_spectrum}. The model offers predictions for scaling in the perpendicular, parallel, and fluctuation directions:

For the perpendicular direction:

\begin{equation}
\zeta^{MS17}_{\lambda}(n)=1-\beta^{n},
\label{eq:11}
\end{equation}

for the parallel direction:

\begin{equation}
\zeta^{MS17}_{\ell_{||}}(n)= 2(1-\beta^{n}),
\label{eq:12}
\end{equation}

and for the fluctuation direction:

\begin{equation}
\zeta^{MS17}_{\xi}(n)= \frac{n(1-\beta^{n})}{n/2 +1 -\beta^{n}},
\label{eq:13}
\end{equation}

where $\beta = 1/\sqrt{2}$. A more practical discussion of the model is presented in \cite{Schekochihin_2022}.

\section{Data Analysis}\label{sec:data analysis }

To investigate and quantify the three-dimensional anisotropy of higher-order magnetic field moments, we employ a methodology proposed by \cite{Wang_2022}, which builds upon and extends the framework established in \cite{Chen_2012ApJ}.

\par Adhering to the approach outlined in \cite{Chen_2012ApJ}, we establish a locally-defined, scale-dependent Cartesian coordinate system, represented as ($\hat{\xi}$, $\hat{\lambda}$, $\hat{\ell_{||}}$). In this coordinate system the ``parallel'' direction, $\hat{\ell_{||}}$, is aligned with the local magnetic field, $\boldsymbol{B_{\ell}}$, defined by:

\begin{equation}
\boldsymbol{B_{\ell}} = [\boldsymbol{B}(\boldsymbol{r} + \boldsymbol{\ell}) + \boldsymbol{B}(\boldsymbol{r})]/2,
\label{eq:15}
\end{equation}

where $\boldsymbol{\ell} = \tau  (\boldsymbol{V}_{\ell} - \boldsymbol{V}_{\text{sc}})$ represents the displacement vector, where $\boldsymbol{V}_{\ell}$ and $\boldsymbol{V}_{\text{sc}}$ are the local scale-dependent velocity field and the spacecraft velocity, respectively, in the RTN coordinate system \citep{FRANZ2002217}.
Magnetic field increments are calculated using:

\begin{equation}
\delta \boldsymbol{b} = \boldsymbol{B}(\boldsymbol{r} + \boldsymbol{\ell}) - \boldsymbol{B}(\boldsymbol{r}).
\label{eq:14}
\end{equation}

 The amplitude of the field increment is denoted as $\delta b = |\delta \boldsymbol{b}|$. The local ``displacement'' direction, $\hat{\xi}$, aligns with the unit vector of the perpendicular component of the field increment, with $\delta \boldsymbol{b}_{\perp}$ defined as:

\begin{equation}
\delta \boldsymbol{b}_{\perp} =  \boldsymbol{B_{\ell}} \times (\delta \boldsymbol{b} \times \boldsymbol{B_{\ell}}).
\label{eq:16}
\end{equation}

Lastly, the ``perpendicular'' direction, $\hat{\lambda}$, is orthogonal to both $\hat{\xi}$ and $\hat{\ell_{||}}$, $\hat{\lambda} = \hat{\ell_{||}} \times \hat{\xi} $. The Cartesian system can be converted into a spherical polar coordinate system ($\ell$, $\theta_{B}$, $\phi_{\delta \boldsymbol{B}_{\perp}}$), where $\theta_{B}$ is the angle between $\boldsymbol{B}_{\ell}$ and $\boldsymbol{\ell}$, and $\phi_{\delta \boldsymbol{B}_{\perp}}$ is the angle between $\hat{\xi}$ and the projection of $\boldsymbol{\ell}$ onto the plane orthogonal to $\boldsymbol{B}_{\ell}$.

\par In our analysis, we utilize the 5-point (5-point) increment method, which represents a significant advancement over the conventional 2-point (2-point) method, especially for examining turbulence statistics in sub-ion regimes \citep{2019_Cerri}. A critical benefit of the 5-point method is its reduced susceptibility to large-scale spectral leakage \citep{2019ApJ...874...75C}, rendering it more suitable and effective for our ensuing analysis.

\par For calculating 5-point structure functions, denoted as $SF^{n}_{5}$, a modified definition of $\delta \boldsymbol{b}$ is required 

% \begin{gather}
%     \delta \boldsymbol{b} = \left[\boldsymbol{B}(\boldsymbol{r}-2\boldsymbol{\ell})-4 \boldsymbol{B}(\boldsymbol{r}-\boldsymbol{\ell})+6 \boldsymbol{B}(\boldsymbol{r}) \right.\nonumber\\
%     \phantom{\delta \boldsymbol{b} = \left[\right.} \left.-4\boldsymbol{B}(\boldsymbol{r}+\boldsymbol{\ell})+ \boldsymbol{B}(\boldsymbol{r}+2\boldsymbol{\ell})\right]/\sqrt{35}. \label{eq:17}
% \end{gather}

\begin{eqnarray}
    \delta \boldsymbol{b} &=& \frac{1}{\sqrt{35}}\left[\boldsymbol{B}(\boldsymbol{r}-2\boldsymbol{\ell})-4 \boldsymbol{B}(\boldsymbol{r}-\boldsymbol{\ell}) \right. \nonumber \\
    && \left.+6 \boldsymbol{B}(\boldsymbol{r}) -4\boldsymbol{B}(\boldsymbol{r}+\boldsymbol{\ell})+ \boldsymbol{B}(\boldsymbol{r}+2\boldsymbol{\ell})\right].
\label{eq:17}
\end{eqnarray}
 Moreover, the local scale-dependent value $\boldsymbol{\psi}_{\ell}$ of a field $\boldsymbol{\psi}$ can be computed as a weighted average using a five-point stencil:
% \begin{multline}
% \boldsymbol{\psi}_{\ell} = \left[\boldsymbol{\psi}(\boldsymbol{r} - 2\ell) + 4\boldsymbol{\psi}(\boldsymbol{r} - \ell) + 6\boldsymbol{\psi}(\boldsymbol{r}) \right.\\
% \left.+ 4\boldsymbol{\psi}(\boldsymbol{r} + \ell) + \boldsymbol{\psi}(\boldsymbol{r} + 2\ell)\right]/16.
% \label{eq:18}
% \end{multline}

\begin{eqnarray}
\boldsymbol{\psi}_{\ell} &=& \frac{1}{16}\left[\boldsymbol{\psi}(\boldsymbol{r} - 2\ell) + 4\boldsymbol{\psi}(\boldsymbol{r} - \ell) \right. \nonumber \\
&& \left. + 6\boldsymbol{\psi}(\boldsymbol{r})+ 4\boldsymbol{\psi}(\boldsymbol{r} + \ell) + \boldsymbol{\psi}(\boldsymbol{r} + 2\ell)\right].
\label{eq:18}
\end{eqnarray}

For example, the local scale-dependent magnetic and velocity fields are represented by $\boldsymbol{B}_{\ell}$ and $\boldsymbol{V}_{\ell}$, respectively. 

\par The $\mathit{n}$th-order, structure functions conditioned on the pair of angles $\theta_{B}, ~ \phi_{\delta \boldsymbol{B}_{\perp}}$, are defined as:
 \begin{equation}
            SF^{n}(\ell, ~ \theta_{B}, ~ \phi_{\delta \boldsymbol{B}_{\perp}}) = \langle (\delta B)^{n}|~\theta_{B}, ~ \phi_{\delta \boldsymbol{B}_{\perp}}, ~\ell \rangle,
      \label{eq:19}
 \end{equation}

The conditional average in Equation \ref{eq:19} was calculated over the angle bin $\omega (i-1)^{\circ} \leq \theta_{B} \leq \omega i^{\circ}$, $ \omega (j-1)^{\circ} \leq \phi_{\delta \boldsymbol{B}_{\perp}} \leq \omega j^{\circ}$, where $i =1, ..., 9$  and $j=1, ..., 9$.  In the following, $\omega$ takes the value of $\omega = 5$ for estimating lower-order moments and $\omega = 10$ for higher-order moments.

In the following, we focus mainly on three special cases, defining the components in 
\begin{equation}
     i=1,\quad j=1:  \quad SF(\ell_{||})^{n}, \quad 
 \quad \text{``parallel''}, \quad \label{eq:20}
\end{equation}
\begin{equation}
     i=9,\quad j=1: \quad SF(\xi)^{n}, \quad \text{``displacement''}, \label{eq:21}
\end{equation}
\begin{equation}
     i=9,\quad j=9: \quad SF(\lambda)^{n}, \quad \text{``perpendicular''}, \label{eq:22}
\end{equation}

directions, where  $ \ell_{||} = \boldsymbol{\ell} \cdot \boldsymbol{\hat{\ell}_{||}} , ~ \lambda = \boldsymbol{\ell} \cdot \boldsymbol{\hat{\lambda}}$, and $\xi = \boldsymbol{\ell} \cdot \boldsymbol{\hat{\xi}}$. 

\par To estimate a component structure function for the entire dataset, we adopt the methodology outlined in \cite{Verdini_3D_2018}. For each selected interval, $j$, within the dataset, we compute the structure functions $SF^{n}_{j}(\ell, ~\theta_{B}, ~\phi_{\delta \boldsymbol{B}_{\perp}})$ for the three orthogonal components, as defined by Equations \ref{eq:20} to \ref{eq:22}. Considering the substantial variation in the root mean square (rms) of fluctuations between intervals, normalization is a critical step prior to averaging these intervals. The normalization involves selecting a specific scale, $\ell_{\ast}$, and normalizing each $SF^{n}_j$ by the energy of fluctuations at that scale. We determine an appropriate $\ell_{\ast}$ by estimating the trace structure function $S^{n}_j(\ell)$ for each interval and identifying a scale range where power-law behavior is consistent across all $S^{n}_j(\ell)$. The fluctuation energy at scale $\ell_{\ast}$ is given by the value of the trace structure function $S_i(\ell_{\ast})$.

The normalized weighted average structure function for a given magnetic field component is then calculated as follows:

\begin{equation}
\Tilde{SF}^{n}(\ell,\theta_{B}, \phi_{\delta \boldsymbol{B}_{\perp}}) = \sum_{j} \frac{SF^{n}_{j}(\ell,\theta_{B}, \phi_{\delta \boldsymbol{B}_{\perp}})}{S^{n}_{j}(\ell_{\ast})} W_{j},
\label{eq:23}
\end{equation}

\begin{figure*}
     \centering
     \includegraphics[width=1\textwidth]{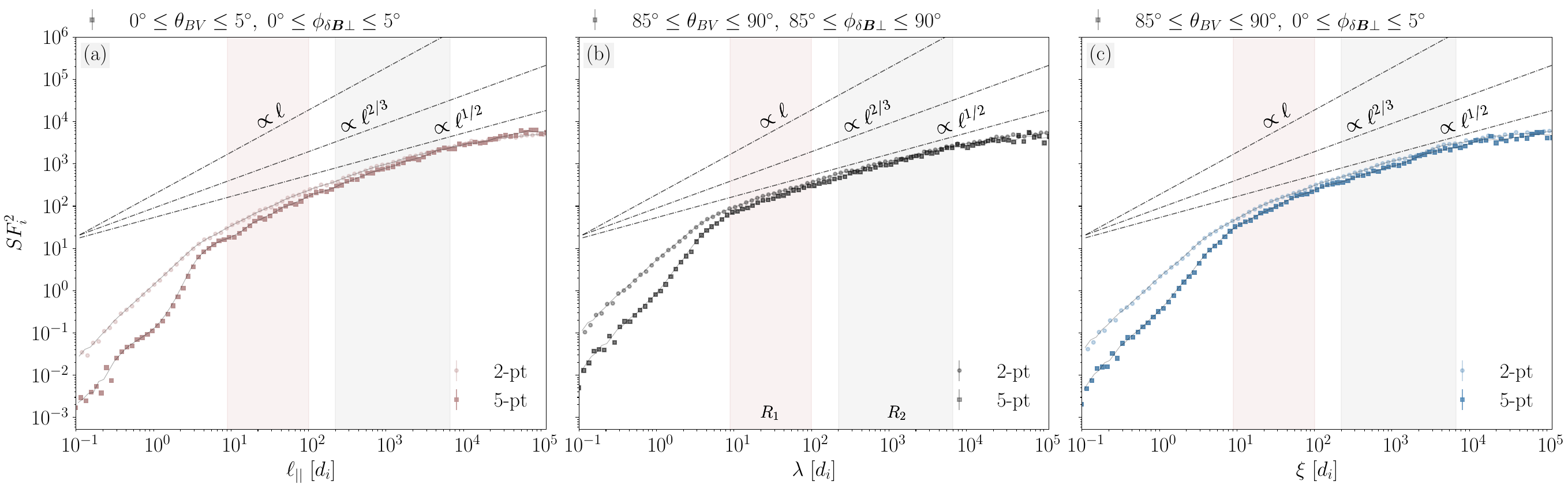}
      % \includesvg[width=1\textwidth]{figures/fig_2pt_5pt_sfuncs_comparison.pdf}
     \caption{The local 2-point structure functions (circles) and 5-point structure functions (squares) averaged for the five most highly Alfv\'enic intervals within our dataset. The structure functions are displayed for the parallel, perpendicular, and displacement directions, indicated by red, black, and blue colors, respectively. Reference lines representing scalings of 1/2, 2/3, and 1 are included for comparison.}
     \label{fig:1_2pt_5pt_comparison}
\end{figure*}

\begin{figure*}
     \centering
     \includegraphics[width=1\textwidth]{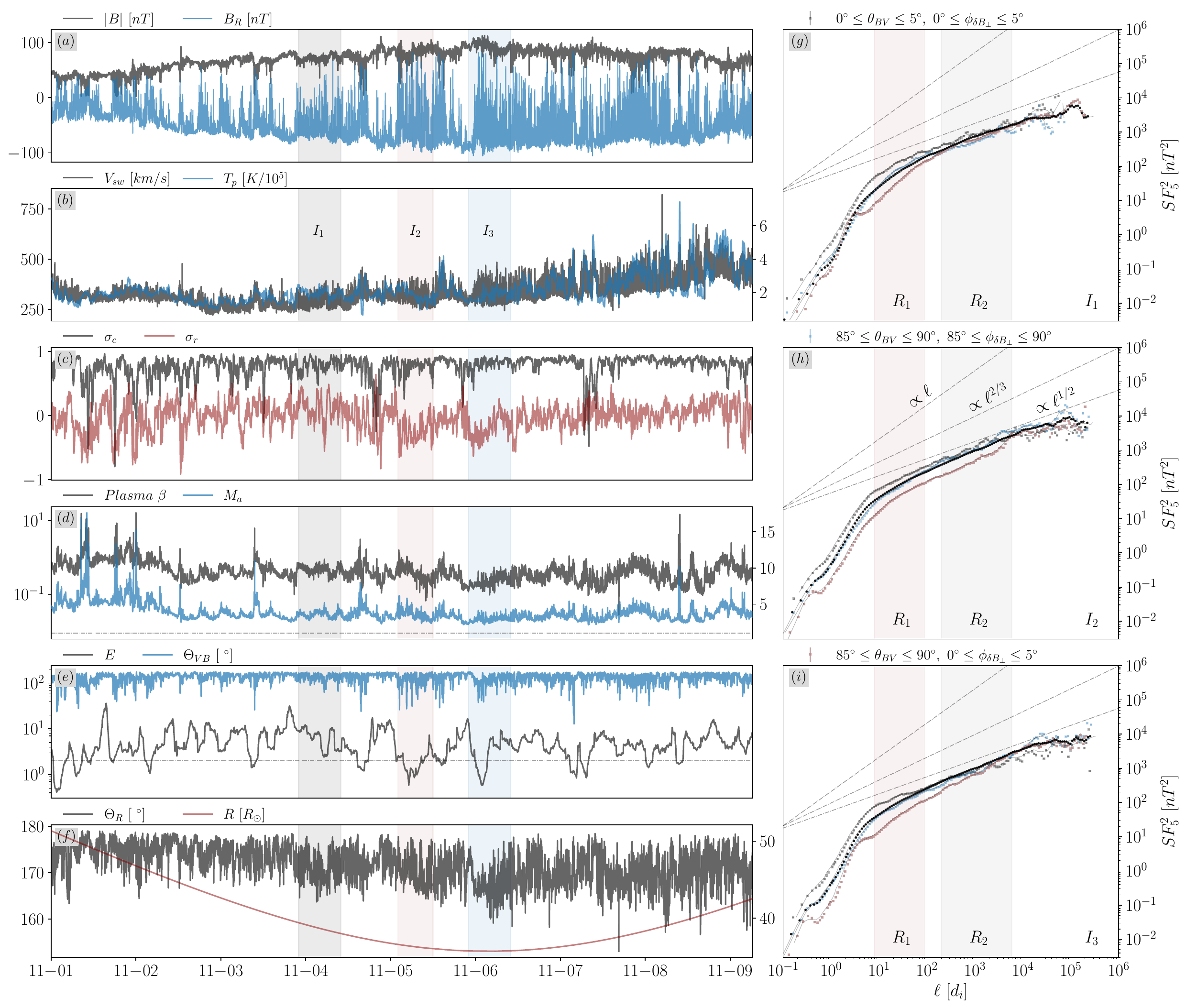}
      % \includesvg[width=1\textwidth]{figures/fig_2pt_5pt_sfuncs_comparison.pdf}
     \caption{Overview of E1: (a) Magnetic field timeseries, radial component, $B_R$ (blue) and magnitude, $|B|$ (black); (b) Solar wind speed, $V_{sw}$ (black, left axis) and proton temperature, $T_p$ (blue, right axis); (c) Normalized cross helicity, $\sigma_c$ (black) and normalized residual energy, $\sigma_r$ (red); (d) Plasma $\beta$ (black, left axis) and Alfv\'enic Mach number, $M_{a} = V_{R}/ |V_{a}|$ (blue, right axis); (e) Variance anisotropy, $E = (b^{2}_{T} + b^{2}_{N})/b^{2}_{R}$, where $b$ represents the rms amplitude of fluctuations (black, left axis) and angle between the magnetic field and velocity flow, $\Theta_{VB}$ (blue); (f) Sampling angle, $\Theta_{R}$, defined as the angle between $\hat{R}$ and $\boldsymbol{V}_{sc} - \boldsymbol{V}_{sw}$ (black, left axis), and radial distance from the Sun, $R$ (red, right axis). Additionally, three intervals denoted as $I_{j}, ~j=~1,~2,~3$ and marked with black, pink, and cyan shadings on the main figure. The corresponding 5-point structure functions of the parallel, perpendicular, and displacement directions, denoted by red, transparent black, and blue colors, respectively, are shown in panels (g)-(i). In addition, the trace structure function is shown in black circles. Reference lines representing scalings of 1/2, 2/3, and 1 are included for comparison.}
     \label{fig:overview}
\end{figure*}

where $W_{j} =n_{j}(\ell, ~\theta_{B}, ~\phi_{\delta \boldsymbol{B}_{\perp}})/n(\ell, ~\theta_{B}, ~\phi_{\delta \boldsymbol{B}_{\perp}})$ represents the weighting factor, with $n_{j}(\ell,~ \theta_{B}, ~\phi_{\delta \boldsymbol{B}_{\perp}})$ being the total number of measurements within each bin for the interval under consideration, divided
by the count in each bin for the whole data set, $n =\sum_{j} n_{j} $.

Our subsequent analysis, as discussed in Section \ref{sec:Data}, relies on these conditionally defined structure functions, utilizing data from the first perihelion of the Parker Solar Probe (PSP). Unless specified otherwise, the results in the following analysis are derived from estimating 5-point structure functions.\footnote{The algorithm detailed in this section, along with a package for downloading, cleaning, and processing PSP data, is readily accessible in \href{https://github.com/nsioulas/MHDTurbPy}{\textcolor{bluegray}{MHDTurbPy}} \citep{MHDTurbPy_Sioulas}.}

\section{Data Set} \label{sec:Data}

We analyze magnetic field and particle data collected during the first perihelion of the PSP mission, covering the period from November 1 to November 11, 2018. Specifically, we analyze magnetic field measurements obtained by the FIELDS instrument \citep{bale_fields_2016}. In particular, we make use of the SCaM data product, which combines measurements from fluxgate and search-coil magnetometers (SCM) by using frequency-dependent merging coefficients. This approach allowed us to observe the magnetic field over a frequency range ranging from direct current (DC) to 1 MHz while achieving optimal signal-to-noise ratio \citep{scam_bowen}. The FIELDS magnetometer suite is susceptible to narrow-band coherent noise stemming from the spacecraft reaction wheels, including their rotation frequencies, as well as harmonic and beat frequencies.  To address potential contamination of magnetic field measurements at ion-scales by the reaction wheels, for each interval, we implement a procedure involving the identification and elimination of reaction wheel noise using the method detailed in \citet{Shankarappa_2023}.

\par Moreover, we incorporated data from the Solar Probe Cup (SPC) instrument, which is part of the Solar Wind Electrons Alphas and Protons (SWEAP) suite \citep{kasper_solar_2016}, to estimate bulk plasma properties. We also utilized Quasi Thermal Noise (QTN) electron density measurements \citep{moncuquet_first_2020,pulupa_solar_2017}. To enhance statistical robustness and expand the sample size, the data were segmented into 12-hour intervals. These intervals were designed to overlap by 6 hours to maximize data utilization. We then conditioned the intervals based on $\sigma_c$, selecting only those with an average value of $\sigma_{c}(\ell_{\ast}) \geq 0.75$, where $\ell_{\ast} =10^{4} d_i$. This methodology yielded a total of 82 intervals, sampled at distances ranging from 0.166 to 0.244 au.

\begin{table}%[htb]
\centering
\caption{ The median values for the spectral indices of the trace, parallel, perpendicular, and displacement components of the magnetic field in the ranges $R_1$ and $R_2$. These indices are derived from the corresponding scaling indices of $SF_{5,~i}^{2}$, utilizing the relationship $\alpha_{i} = -1-\beta_{i}$ \citep{monin_statistical_1987}. The error values provided represent the standard deviation of the mean.}

\vspace{1em}
\resizebox{\columnwidth}{!}{%
\begin{tabular}{c*{5}{>{$}c<{$}}}
\toprule % Top horizontal line
\toprule % Top horizontal line
   & \text{$\alpha$} & \text{$\alpha_{\ell_{||}}$} & \text{$\alpha_{\lambda}$} & \text{$\alpha_{\xi}$}     \\
\midrule 
$R_1$   & -1.79\pm 0.06  & -1.97 \pm 0.05  & -1.64\pm 0.04 & -1.94 \pm 0.06   \\
$R_2$   & -1.53\pm 0.02  & -1.66 \pm 0.05  & -1.49\pm 0.03 & -1.56\pm 0.08   \\
\bottomrule % Bottom horizontal line
\end{tabular}
}
\end{table}\label{tab:1}

\begin{figure*}
     \centering
     \includegraphics[width=1\textwidth]{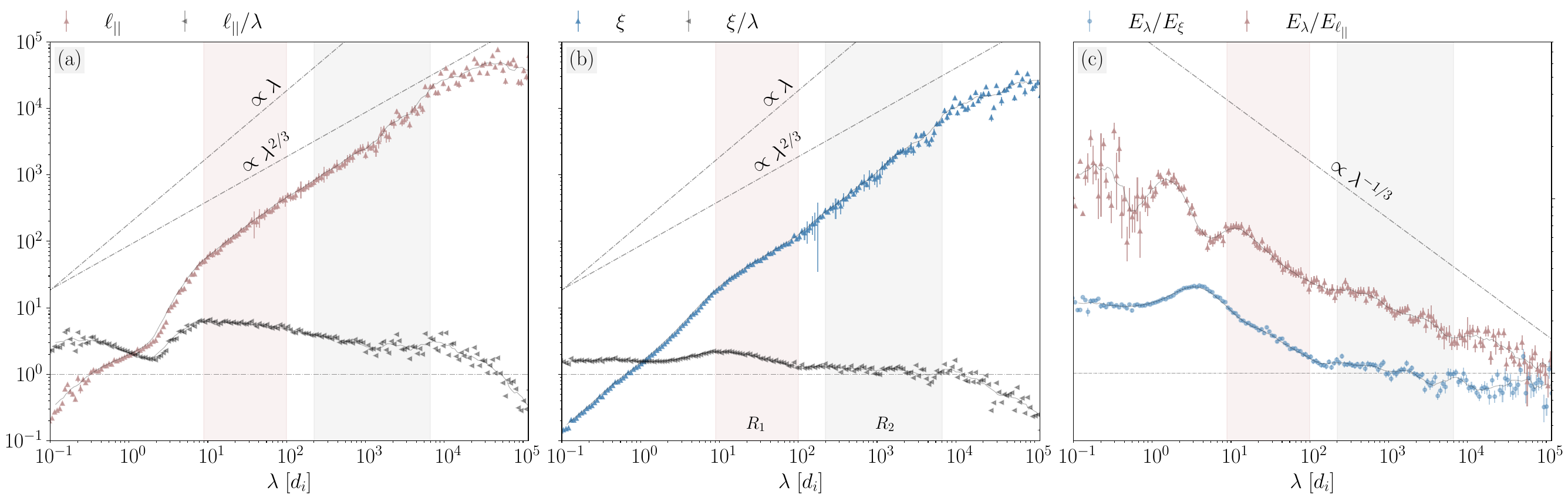}
     \caption{Wavevector anisotropy relationships: (a) $\ell_{||} = \ell_{||}(\lambda)$ (b) $\xi = \xi(\lambda)$,  determined by equating pairs of $SF^{2}_{5}(\lambda)$ with $SF^{2}_{5}( \ell_{||})$, and $SF^{2}_{5}(\lambda)$ with $SF^{2}_{5}(\xi)$, respectively. The gray lines show the aspect ratios, $\ell_{||}/\lambda$ and $\xi/\lambda$ plotted against $\lambda$ in panels (a) and (b), respectively. For context, reference lines indicating scalings of 2/3 and 1 are also included. Panel (c) presents power anisotropy, with $E_{\lambda}/E_{\ell_{||}}$ depicted in red and $E_{\lambda}/E_{\xi}$ in blue. }
     \label{fig:wavevector_anisotropy}
\end{figure*}

\section{Results} \label{sec:Statistical Results}

\subsection{A Comparison of \texorpdfstring{$SF_{5}^{n}$ and $SF_{2}^{n}$}{SF5^n and SF2^n}}\label{subsec:Sf2_Sf5}

This section is dedicated to a comparative analysis aimed at substantiating our preference for the 5-point structure function method over the traditionally favored 2-point approach. To this end, we calculated second-order structure functions for the parallel, perpendicular, and displacement components according to Equations \ref{eq:20} to \ref{eq:22}, setting $\omega = 5$. This involves estimating $\boldsymbol{B}_{\ell}$ and $\delta \boldsymbol{b}$ as per Equations \ref{eq:15} to \ref{eq:16} for $SF_{2}^{2}$, and Equations \ref{eq:18} to \ref{eq:19} for $SF_{5}^{2}$.

The local 2-point structure functions (circles) and 5-point structure functions (squares) averaged for the five most highly Alfv\'enic intervals within our dataset are illustrated in Figure \ref{fig:1_2pt_5pt_comparison}. To highlight specific scale ranges, pink and gray shadings are employed for intervals $8-100d_i$ (labeled $R_1$) and $200-6000d_i$ (labeled $R_2$), respectively.  At large scales, comparable results are obtained from both the 5-point and 2-point methods. However, a marked divergence is observed in the $R_1$ range. The 5-point method reveals steep scaling for parallel and displacement components, with indices $\beta_{\ell_{||}} \approx \beta_{\xi} \approx 1$, aligning with wavelet-derived parallel component scaling of the same dataset reported in \cite{Sioulas_2023_anisotropic}. In contrast, the 2-point method produces a notably flatter slope. 

The distinction between the two methods becomes stark at kinetic scales, highlighting the 2-point method's limitations in detecting steep scalings. This shortfall is further evident when comparing $\boldsymbol{B}$-trace wavelet structure functions using both the $SF_{2}^{2}$ and $SF_{5}^{2}$ methods; only the 5-point approach produces scalings that align with wavelet analyses across scales from injection to kinetic (details not shown here). This inconsistency underscores potential inaccuracies when employing the 2-point method in scenarios characterized by steep scaling, as also emphasized by \cite{2019_Cerri}\footnote{A cautionary note is warranted: when comparing $SF_{5}^{2}$ with wavelet-derived trace structure functions for intervals observed in later PSP encounters —characterized by shallower than $1/f$ energy injection scale power spectra \citep{2023_Huang, Davis_2023}— $SF_{5}^{2}$ fails to replicate the wavelet scalings. This indicates that $SF_{5}^{2}$, similar to its $SF_{2}^{2}$ counterpart, is ineffective under conditions with scalings shallower than $1/f$.}. Therefore, for our further analyses, we have chosen to rely on the 5-point method. 

\par Across all examined intervals, the results remained qualitatively consistent. Three such intervals are illustrated in panels (g)-(h) of Figure \ref{fig:overview}. In addition to the component structure functions, these panels also illustrate the trace structure function estimated for the respective intervals. It can be observed that the perpendicular component can significantly diverge from the trace, while the latter typically shows remarkable overlap with the displacement component, with the difference becoming more pronounced towards smaller scales. 

Table \ref{tab:1} presents the median $SF_{5}^{2}$ scalings estimated across the $R_1$ and $R_2$ ranges for the entire dataset.

\subsection{Power \& wavevector anisotropy }\label{subsec:anisotropy}

To examine the scale-dependent three-dimensional anisotropy in our dataset, we calculated $SF_{5}^{2}(\ell)$, following the methodology outlined in Equations \ref{eq:20} to \ref{eq:22}, setting $\omega = 5$. Our analysis commenced by identifying the anisotropic relationships for individual intervals, then proceeded to compute a scale-dependent median for the entire dataset, utilizing 150 logarithmically spaced bins. 

Panels (a) and (b) of  Figure  \ref{fig:wavevector_anisotropy} illustrate the anisotropic relationships  $\ell_{||}(\lambda)$, and $\xi(\lambda)$, derived by equating pairs of structure functions: $SF^{2}_{5}(\lambda)$ with $SF^{2}_{5}(\ell_{||})$, and $SF^{2}_{5}(\lambda)$ with $SF^{2}_{5}(\xi)$, respectively. The aspect ratios $\ell_{||}/\lambda$ and $\xi/\lambda$ are represented by gray lines in their respective panels. Panel (c) focuses on power anisotropy, illustrating the ratios $SF^{2}_{5}(\lambda)/SF^{2}_{5}(\ell_{||})$ and $SF^{2}_{5}(\lambda)/SF^{2}_{5}(\xi)$ in red and blue, respectively. The median values of the dataset's anisotropic scalings are summarized in Table \ref{tab:2}.

\par At large scales, our observations reveal a rough equipartition of the fluctuating energy between $SF^{2}_{5}(\lambda)$ and $SF^{2}_{5}(\ell_{||})$, with $SF^{2}_{5}(\xi)$ being slightly more energetic. The energy distribution is reflected in the wavevector anisotropy and aspect ratios, indicating that eddies tend to be slightly compressed along the fluctuation direction. 

\begin{table}%[htb]
\centering
\caption{  Median values of the scaling indices for wavevector anisotropies, specifically $\ell_{||} \propto \lambda^{w_{\ell_{||}}}$ and $\xi \propto \lambda^{w_{\xi}}$, and the power anisotropies $E_{\lambda}/E_{\ell_{||}} \propto \lambda^{p_{\ell_{||}}}$ and $E_{\lambda}/E_{\xi} \propto \lambda^{p_{\xi}}$. These indices were derived by applying a power-law fit to the curves obtained from individual intervals over the scale ranges $R_1$ and $R_2$. The table presents the median values along with their associated errors, represented as the standard deviation of the mean.}

\vspace{1em}
\resizebox{\columnwidth}{!}{%
\begin{tabular}{c*{5}{>{$}c<{$}}}
\toprule % Top horizontal line
\toprule % Top horizontal line
   & \text{$w_{\ell_{||}}$} & \text{$w_{\xi}$} & \text{$p_{\ell_{||}}$} & \text{$p_{\xi}$}     \\
\midrule 
$R_1$   & 0.89\pm 0.06  &  0.77 \pm 0.05  & -0.31\pm 0.04 & -0.30 \pm 0.07   \\
$R_2$   & 0.86\pm 0.08  &  0.99 \pm 0.06  & -0.19\pm 0.05 & -0.09\pm 0.06   \\
\bottomrule % Bottom horizontal line
\end{tabular}
}
\end{table}\label{tab:2}

\begin{figure*}
     \centering
     \includegraphics[width=0.95\textwidth]{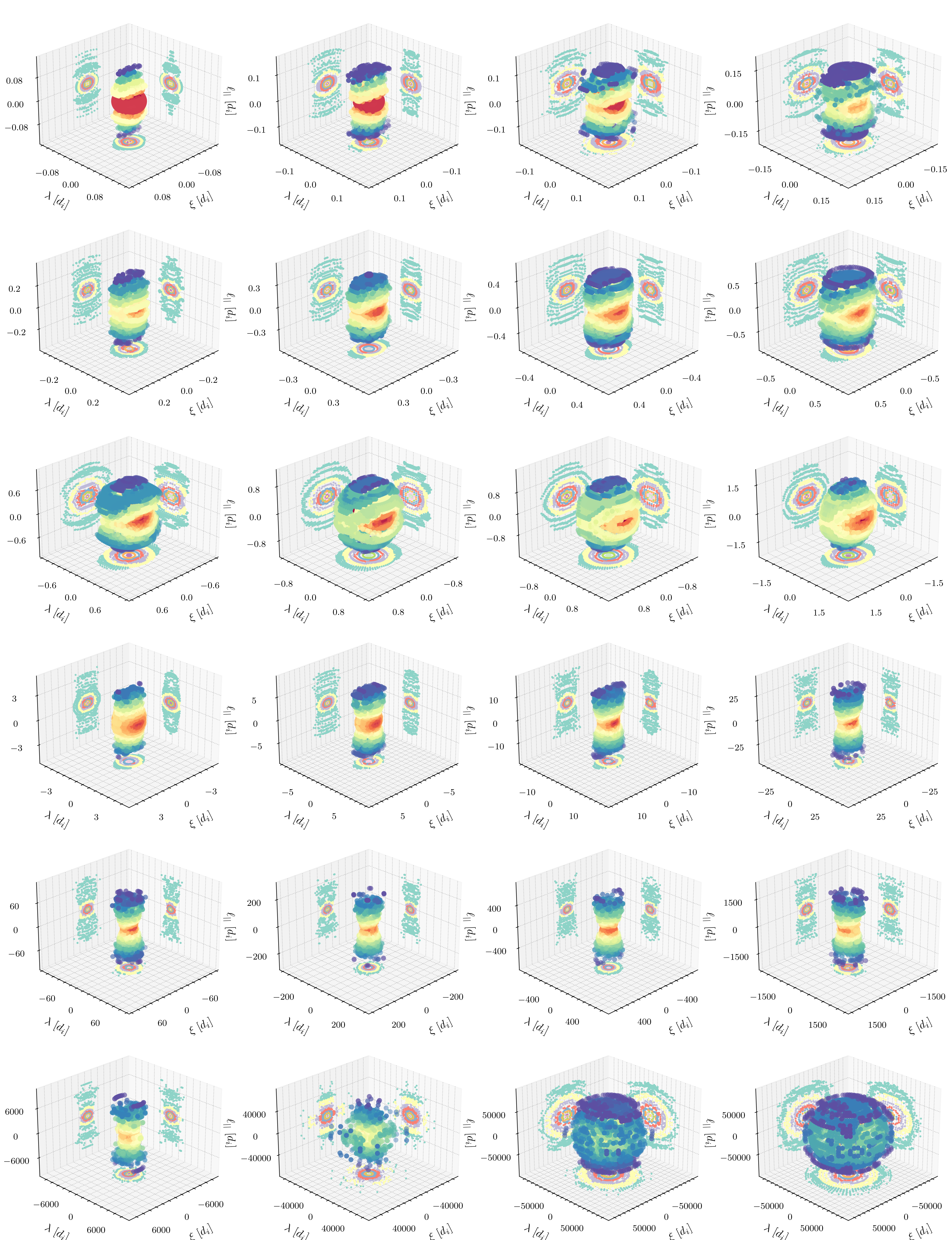}
     \caption{3D representation of turbulence eddies obtained by estimating isosurfaces of constant energy of $SF^{2}_{5}$ at different levels, ranging from small scales (top left) to large scales (bottom right). The color scheme, although redundant, indicates the distance from the origin (0,0,0). Additionally, projections of the object onto each respective plane are displayed. While the projections share a common colormap to denote the distance from each plane's origin, the colormap for the 3D object differs.}
     \label{fig:3D_eddies}
\end{figure*}

Within the $R_2$ range, we note that the fluctuating magnetic energy is distributed almost evenly between the perpendicular and displacement components. This observation suggests rough axisymmetry of the turbulent eddies at these scales, i.e., approximate isotropy in the plane perpendicular to $\boldsymbol{B}_{\ell}$. The aspect ratio $\xi/\lambda$ is observed to be close to, yet marginally greater than, unity, displaying only a slight increase within $R_2$. This trend is further emphasized by the scale-dependent power-anisotropy depicted in panel (c). In contrast, within this range, eddies exhibit elongation along $\boldsymbol{B}_{\ell}$, as indicated by the ratio $\ell_{||}/\lambda$, which is greater than 1 and shows a monotonic increase towards smaller scales. These findings collectively indicate that turbulent eddies within the $R_2$ range predominantly exhibit a field-aligned tube topology, consistent with the results presented in \cite{vinogradov2023embedded}. This is visually demonstrated in Figure \ref{fig:3D_eddies}.
At scales smaller than $\lambda \lesssim 100 d_i$, a noticeable shift from isotropy in the plane perpendicular to $\boldsymbol{B}_{\ell}$ becomes evident. This shift is highlighted by a gradual increase in the aspect ratio $\xi/\lambda$, indicating a transition of eddy structures from tube-like to ribbon-like. The evolving eddies exhibit three-dimensional anisotropy, adhering to the relationship $\ell_{||} \gg \xi \gg 
\lambda$. This trend persists into the smaller-scale end of the $R_1$ range, where both power-anisotropy ratios follow a scaling of approximately -1/3. Additionally, within $R_1$, the ratio $\ell_{||}/\lambda$ continues to rise, albeit at a reduced rate compared to the $100-600 d_i$ range, where $\ell_{||} \propto \lambda^{0.72 \pm 0.04}$. This slower rate of increase within $R_1$ can be attributed to the steepening of $SF^{2}_{5}(\ell_{||})$ at scales marginally larger than $R_1$.

As we move below the $R_1$ range and into the transition region \citep{Sahraoui_2009_transition, Bowen_2020_transition}, the previously observed trend of increasing anisotropy ceases. Within the scale range of $2 d_i \lesssim \lambda \lesssim 8 d_i$, the eddies start to demonstrate more isotropic characteristics. This tendency towards isotropy peaks at $\lambda = 2 d_i$, where the aspect ratio reaches $\lambda:\xi:\ell = 1.56:1:1$. Throughout this scale range, the anisotropic scaling relations ---derived from fitting curves to individual intervals and estimating the median values, presented along with the standard deviation of the mean--- conform to $\ell_{||} \propto \lambda^{2.01 \pm 0.06}$ and $\xi \propto \lambda^{1.25 \pm 0.05}$.

At even smaller scales, distinct scaling anisotropies characterize two separate ranges. Within $1 d_i \lesssim \lambda \lesssim 2 d_i$, the ratio $\ell_{||}/\lambda$ shows an upward trend, following $\ell_{||} \propto \lambda^{0.67 \pm 0.02}$, while the ratio $\xi/\lambda$ remains relatively stable, adhering to $\xi \propto \lambda^{0.98 \pm 0.04}$. These findings stand in contrast to standard kinetic Alfv\'en wave (KAW) turbulence models \citep{Howes_2008, Schekochihin_2009_review} and deviate statistically from in-situ observations reported in \citep{Duan_2021, Zhang_2022}. However, they are in agreement with the intermittent KAW model proposed by \citep{Boldyrev_2012_kinetic} and align with numerical kinetic simulations by \citep{2019_Cerri}, as well as in-situ observations in the magnetosheath \citep{Wang_2020}. At yet smaller scales, $0.5 d_i \leq \lambda \leq 1 d_i$, both $\ell_{||}/\lambda$ and $\xi/\lambda$ ratios exhibit an increase, scaling as $\ell_{||} \propto \lambda^{0.5 \pm 0.05}$ and $\xi \propto \lambda^{0.87 \pm 0.04}$, respectively.

 Figure \ref{fig:3D_eddies} presents a three-dimensional representation of turbulence eddies, illustrating isosurfaces of $SF^{5}_{2}$ at various scales. This visualization was achieved by estimating conditional $SF^{2}_{5}(\ell, ~\theta_{B}, ~ \phi_{\delta \boldsymbol{B}_{\perp}})$ , according to Equation \ref{eq:19}, and utilizing $5^{\circ}$ angular bins. The spherical polar coordinates $(\ell, ~\theta_{B}, ~ \phi_{\delta \boldsymbol{B}_{\perp}})$ obtained from this process were then converted into Cartesian coordinates $(\ell_{||}, ~\xi, ~ \lambda)$. Surfaces computed for the first octant were mirrored across to the other octants, based on the assumption of reflectional symmetry \citep{Chen_2012ApJ}. In the 3D visualization, surface colors represent the distance from the origin, with cooler colors indicating larger distances. When these surfaces are projected onto different planes, the color denotes the distance from the origin of each respective plane. It is important to note that the colormap applied to these planar projections differs from the one used for the 3D representation. The color coding in these projections reflects the range of maximum and minimum distances observed across all three components.

\subsection{Higher order statistics \& Intermittency}\label{subsec:SFs}

\par We computed five-point structure functions, $SF_{5}^{n}$, for parallel, perpendicular, and displacement components, as per Equations \ref{eq:20} to \ref{eq:22}, considering orders $n = 1, \ldots, 10$ with $\omega = 10$. Additionally, we evaluated $\boldsymbol{B}$-trace structure functions, $S_{5}^{n}$. Furthering our analysis, we derived thenormalized weighted average, $\Tilde{SF}{5}^{n}$, for our dataset following Equation \ref{eq:23}. After identifying two distinct sub-inertial ranges displaying clear power-law behavior, we calculated two sets of normalized structure functions. Within the $R_1$ domain, we normalized $SF_{5}^{n}$ using $S_{5}^{n}(\lambda_{\ast})$, where $\lambda_{\ast} = 50 d_i$.

 \begin{figure*}
     \centering
     \includegraphics[width=1\textwidth]{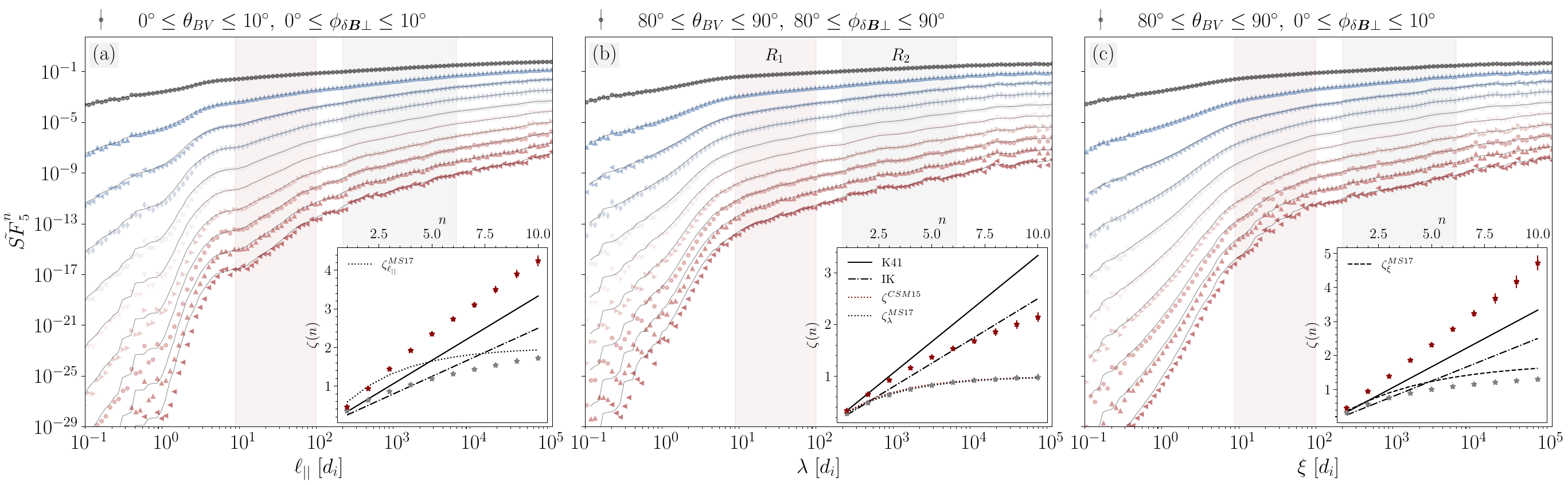}
     \caption{The main panels depict the normalized weighted average structure functions, $\tilde{SF}^{n}_{5}(\ell)$, for (a) parallel, (b) perpendicular, and (c) displacement components, each adjusted with a vertical offset for enhanced clarity. Prior to the weighted average estimation, each $SF^{n}_{5}(\ell)$ was normalized using the value of the trace $S^{n}_{5}(\lambda{\ast})$, where $\lambda_{\ast} = 2 \times 10^{3} d_i$. These normalized averages, $\tilde{SF}^{n}_{5}(\ell)$, displayed in the main figure, were used to estimate $\zeta_{n}$ for $R_2$, marked as gray asterisks in the insets. For $\zeta_{n}$ estimation in $R_1$, indicated by red asterisks in the insets, a similar normalization process was applied at $\lambda_{\ast} = 50 d_i$. It is important to note that the latter normalized $\tilde{SF}^{n}_{5}(\ell)$ for $R_1$  are not depicted in the figure. Error bars in the insets represent the uncertainty associated with the power-law fits. For comparison, the scaling behaviors as predicted by the \citetalias{kolmogorov_local_1941}, \citetalias{iroshnikov_turbulence_1963}, \citetalias{chandran_intermittency_2015}, and \citetalias{Mallet_2017} models are also included.}
     \label{fig:sfuncs}
\end{figure*}

For the $R_2$ domain, normalization employed a different scale, $\lambda_{\ast} = 2 \times 10^{3} d_i$. The latter normalized structure functions, particularly for the parallel, perpendicular, and displacement components, are depicted in Figure \ref{fig:sfuncs} panels (a) to (c), respectively. To aid visualization, each n-th order moment was vertically offset by $10^{-n}$. Fitting each component of $\Tilde{SF}_{5}^{n}(\ell)$ to a power law, $\propto \ell^{\zeta_{n}}$, facilitated the estimation of scaling exponents, $\zeta_{n}$. The resulting $\zeta_{n}$ are depicted by red asterisks for the $R_{1}$ domain and gray for $R_{2}$ within the insets of the corresponding figure panels. For comparison, scaling exponent predictions based on the theoretical models proposed by \citetalias{chandran_intermittency_2015} and \citetalias{Mallet_2017} are also included.

\par In the $R_{2}$ range, the scaling exponent $\zeta_n^{\lambda}$ of the perpendicular component forms a convex function of n, indicative of multifractal statistics and strong intermittency. This scaling profile closely aligns with the theoretical predictions by \citetalias{chandran_intermittency_2015} and \citetalias{Mallet_2017}, showing notable correspondence to the latter model at lower $n$ values. For $SF^{n}_{5}(\xi)$, the observed scaling exponents, $\zeta_n^{\xi}$, slightly deviate towards shallower gradients compared to the \citetalias{Mallet_2017} model. The scaling exponents of the parallel component $\zeta_n^{\ell_{||}}$ exhibit a nonlinear dependence on $n$, though with less pronounced concavity than the perpendicular components, and notably diverge from the \citetalias{Mallet_2017} model, even at lower $n$ values. A comprehensive discussion of these findings and their broader implications is provided in Section \ref{sec:Disussion}.

\par In the small-scale sub-inertial range, $R_1$, the scaling exponents for both the parallel and displacement components display a linear relationship with $n$. Conversely, the scaling exponent $\zeta_n^{\lambda}$ of the perpendicular component forms a convex function of n, albeit demonstrating a lesser extent of non-linearity relative to that observed in the $R_2$ range. Overall, $R_1$ is characterized by less pronounced intermittency signatures compared to $R_2$, with the observed $\zeta_n$ profiles deviating from the expectations set by established theoretical models.

\par To further investigate multifractality and deviations from Gaussian statistics in the magnetic field time series, we consider the Scale-Dependent Kurtosis, defined as $K(\ell) = SF^4(\ell) / [SF^2(\ell)]^2$ \citep{frisch_turbulence_1995, bruno_radial_2003}. As a normalized fourth-order moment, $K(\ell)$ is sensitive to extreme values of increments, allowing us to detect the tendency of PDFs in intermittency-affected time series to exhibit increasingly flared-out tails at smaller scales. In simpler terms, it quantifies how the ``tailedness'' of the distribution of increments in a turbulent field changes across various scales.

\par
\begin{figure*}
     \centering
     \includegraphics[width=1\textwidth]{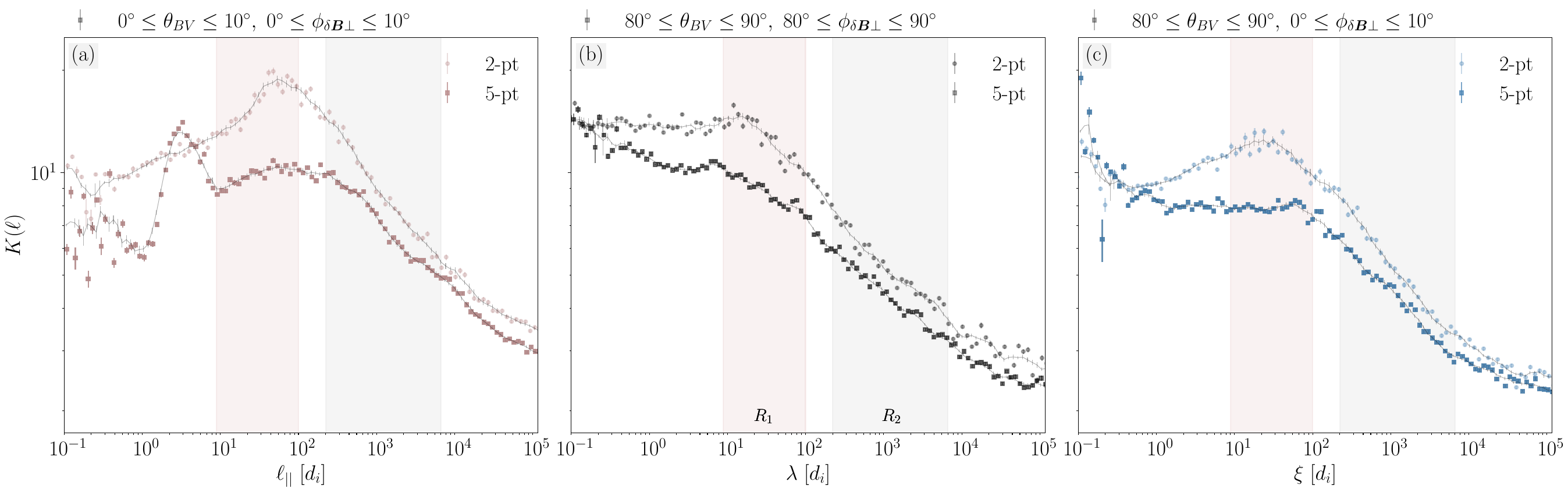}
     \caption{The scale-dependent kurtosis of the magnetic field, denoted as $K_{5}$ and $K_{2}$, is estimated using 5-point structure functions ($SF_{5}^{n}$) and 2-point structure functions ($SF_{2}^{n}$), respectively. These are plotted in red,  gray and light-blue, respectively, as a function of scale in units of ion inertial length ($d_{i}$) for the parallel (a), perpendicular (b), and displacement (c) directions. Power-law fits have been applied to $K_{5}$ over the region $R_{2}$.
     }
     \label{fig:sdk}
\end{figure*}

 We employ both 2-point ($K_{2}(\ell)$) and 5-point ($K_{5}(\ell)$) methods to study the fractal properties of magnetic field time series. The resulting curves for the parallel, perpendicular, and displacement components of the magnetic field are illustrated in panels (a) to (c) of Figure \ref{fig:sdk}. At scales $\lambda \geq 100 d_i$, an increase in $K(\ell)$ is observed for all components towards smaller scales, indicating a progressive deviation from Gaussianity in the underlying PDFs of increments, a hallmark of multifractal statistics \citep{Sorriso-Valvo99}.

\par The limitations of the 2-point method are particularly evident in the $R_1$ range, where it notably diverges from the 5-point approach. In the case of the perpendicular component, $K_{2}(\lambda)$ appears to plateau at $\lambda < 20 d_i$, consistent with the findings of \cite{chhiber_subproton}. Conversely, $K_{5}(\ell)$ maintains an increasing trend in $R_1$, albeit with a less steep slope compared to $R_2$. Within the $R_1$ range, both the parallel, $K_{5}(\ell_{||})$, and  displacement, $K_{5}(\xi)$, components exhibit super-Gaussian but monofractal behavior, consistent with the linear $\zeta_{n}^{\ell_{||}}$ and $\zeta_{n}^{\xi}$ profiles illustrated in Figure \ref{fig:sfuncs}.

\par At kinetic scales, both $K_{5}(\lambda)$ and $K_{5}(\xi)$ exhibit an increasing trend towards smaller scales, while the behavior of the parallel component remains less distinct. This trend in the perpendicular component diverges from the monofractal statistics observed at sub-ion scales in several observational studies using $K_{2}$ \citep{Wu_2013_sdk, Chen_2014, Chhiber_2020}. However, our findings align with hybrid and fully kinetic simulations by \cite{2019_Cerri}, where $K_{5}(\lambda)$ is demonstrated to increase above Gaussian values throughout the sub-ion scale range. 
Furthermore, qualitatively consistent trends were identified by \cite{Alexandrova_2008}, who utilized wavelet-derived kurtosis to observe a gradual increase in this measure at kinetic scales.

\subsection{ Scale-Dependent Dynamic Alignment \& Critical Balance}\label{subsec:result_SDDA}

We begin by examining the scale-dependent behavior of the alignment angle between the perpendicular components of the increments, $\delta \boldsymbol{b}_{\perp} - \delta \boldsymbol{u}_{\perp}$ and $\delta \boldsymbol{z}^{+}_{\perp} - \delta \boldsymbol{z}^{-}_{\perp}$. Els\"asser variable increments were determined as $\delta \boldsymbol{z}_{\perp}^{\pm} = \delta\boldsymbol{v}_{\perp} \pm \text{sign}(B_{r}^{0})\delta \boldsymbol{b}_{\perp}$, where $B_{r}^{0}$ represents the 30-minute rolling average of the radial magnetic field component, $B_{r}$, used to determine the polarity of the background magnetic field. Here, $\boldsymbol{z}_{\perp}^{-}$ and $\boldsymbol{z}_{\perp}^{+}$ denote inward and outward propagating Alfv\'en waves, respectively. Magnetic field data were downsampled, following the application of a low-pass \cite{BUTTERWORTH} filter to mitigate aliasing, to match the temporal resolution of the ion moment data. Magnetic field data were then normalized to velocity units using a 1-minute moving average applied to the proton density, $n_p$, time series.

\par Figure \ref{fig:align_sdda}a,b illustrates two sets of alignment angles: $\theta_{\perp}^{ub(z)}$ in black, as defined by Equation \ref{eq:9}, and $\tilde{\theta}_{\perp}^{ub(z)}$ in blue, calculated according to Equation \ref{eq:10}. These sets are referred to as $\Theta^{z}$ when discussing Els\"asser variables and $\Theta^{ub}$ in the context of velocity-magnetic field fluctuations, with the angle range confined to $0^{\circ}-90^{\circ}$, consistent with \cite{PODESTA_2009_SDDA}. The inset of Figure \ref{fig:align_sdda}b illustrates the normalized residual energy, $\sigma_r$, in red, and the normalized cross helicity, $\sigma_c$, in gray.

\par At the energy injection range, $\lambda \gtrsim 2 \times 10^{4} d_i$, a trend towards tighter alignment at smaller scales is observed, predominantly in $\Theta^{ub}$. This is accompanied by a monotonic increase in $\sigma_c$ and a shift of $\sigma_r$ towards more negative values. These trends are more pronounced over longer observational intervals, although such extended periods are beyond the scope of this analysis.

\par Within the $R_2$ range, $\theta^{z}$ exhibits negligible variation with scale. Conversely, $\tilde{\theta}^{z}$ reveals subtle signatures of enhanced alignment at $\lambda \lesssim 2 \times 10^{3} d_i$, coinciding with $\sigma_r$ transitioning from negative to positive values. Simultaneously, $\Theta^{ub}$ steadily increases, indicating progressive misalignment between the magnetic field and velocity increments at inertial scales.

The disparity between the two alignment definitions becomes more evident in the $R_1$ range, where $\tilde{\theta}^{z} \propto \lambda^{0.11}$. However, observations concerning this range should be approached cautiously due to potential instrumental noise, a matter further explored in Section \ref{sec:Disussion}.

What appears to solidly be the case, however, is that an inverse relationship holds between the alignment angle and the intensity of the field gradients. This relationship is depicted in the inset panel of Figure \ref{fig:align_sdda}a, where $\theta^{ub}$ is plotted across various scale-dependent percentile bins of the Partial Variance of Increments (PVI) diagnostic, $\mathcal{I}_{\boldsymbol{B}}(t , \ell) = |\delta\boldsymbol{B}(t, \ell)|/\sigma_{\boldsymbol{B}}$, with $\sigma_{\boldsymbol{B}}$ representing the standard deviation calculated over a moving window of 1 hour, \citep{greco_partial_2018}. Specifically, at $\lambda \gtrsim 2 \times 10^{2} d_i$, higher-percentile $P(I_{\boldsymbol{B}})$ bins are characterized by lower average $\theta^{ub}_{\perp}$ values. Similar results were obtained when considering  $\theta^{z}_{\perp}$ and segregating alignment angles based on the percentile bins of the PVI diagnostic applied to the $\boldsymbol{z^{+}_{\perp}}$ time-series, $I_{\boldsymbol{z}^{+}}$.

\par Under the assumption that the cascade is local in $\lambda$, we investigated the scale-dependence of the nonlinearity parameter $\chi^{\pm}$, according to the formulations by \citetalias{boldyrev_2006} and \citetalias{chandran_intermittency_2015}: $\chi^{\pm} =  (\ell^{\pm}_{||, \lambda}/ \lambda^{\pm})/(\delta z_{\lambda}^{\mp}/V_{a})sin \theta^{z}$. The analysis of outgoing ($\chi^{+}$) and ingoing ($\chi^{-}$) waves, depicted in gray and red respectively in panel (c) of Figure \ref{fig:align_sdda}, reveals that both cascades start weak, with $\chi^{\pm} < 1$. As the cascade progresses towards smaller scales, a significant increase in the wave to nonlinear times ratio leads to a consistent rise in $\chi^{\pm}$, continuing until scales nearing the $R_2$ range onset.

For the ingoing waves, a transition from weak to strong wave turbulence is noted, with $\chi^{-} > 1$ at $\lambda \approx 3 \times 10^{4} d_i$. The cascade remains strong throughout the resolvable portion of the inertial range, with $\chi^{-}$ being scale-independent, staying close to, yet slightly greater than, 1.

In contrast, the cascade of outwardly propagating waves remains weak within the $R_2$ range. More specifically, $\chi^{+}$ shows a modest increase from approximately 0.1 at $\lambda = 10^{4} d_i$ to around 0.2 at $\lambda = 2 \times 10^{2} d_i$.

The potential inaccuracies in velocity measurements, exacerbated at smaller scales, along with the limited resolution of velocity field data, caution against drawing definitive conclusions about the cascades' nature in the $R_1$ range.

Lastly, it's noteworthy that the definition of $\chi^{\pm}$, as proposed by \citetalias{goldreich_toward_1995}, was also considered. This analysis revealed a scale dependence similar to that of $\chi^{\pm}$, but with both $\chi^{\pm}$ values being approximately twice as high.

% \subsection{Some further diagnostics}\label{subsec:result_furthe_diagnostics}

\section{Discussion}\label{sec:Disussion}

\par Recent in-situ observations indicate that the regime canonically identified as the inertial range comprises two sub-inertial segments, exhibiting distinct scaling behaviors \citep{Wicks_2011, chhiber_subproton, Sioulas_2022_intermittency,telloni_transition, Wu2022OnTS,  Sioulas_2023_anisotropic,Sorriso_Valvo_2023}.

\par Building on this insight, our study investigates the anisotropic properties and higher-order statistics of the two sub-inertial ranges, utilizing a physically motivated, locally defined coordinate system. Concurrently, we focus on evaluating the predictions of homogeneous MHD turbulence models, grounded in the principles of $\it{critical~ balance}$ and $\it{dynamic ~alignment}$, as proposed by \citetalias{chandran_intermittency_2015} and \citetalias{Mallet_2017}. 

\par In the ensuing section, we embark on a detailed comparison with previous theoretical, observational, and numerical results that contextualizes our findings.

\subsection{Investigating the Impact of Imbalance and Expansion on the Higher-Order Statistics}\label{subsec:disc_imbalance_expansion}

\par Phenomenological treatment of homogeneous MHD turbulence (e.g., \citetalias{goldreich_toward_1995}, \citetalias{boldyrev_2006}, \citetalias{chandran_intermittency_2015}, \citetalias{Mallet_2017}) is usually performed under the simplifying assumption of negligible cross helicity. However, the statistical properties of solar wind turbulence vary significantly with the degree of Els\"asser and Alfv\'enic imbalance \citep{Podesta_2010, Chen_2013, Wicks_2011, Wicks_2013_imbalanced_obse, 2018_Bowen_residual, Andres_2019, Sorriso-Valvo_2021, sioulas_22_spectral_evolu, Damicis_2022, McIntyre_2023}. Various models have been proposed as modifications to the frameworks of \citetalias{goldreich_toward_1995} and \citetalias{boldyrev_2006}, incorporating different assumptions about the turbulent cascade to address the imbalance in oppositely directed Alfv\'enic fluxes  \citep{Lithwick_2007_imbalanced_critical_balance,Berenyak_2008, Chandran_2008_cross_hel, PhysRevLett.102.025003, Podesta_Bhattacharjee_2010, Schekochihin_2022}. As these models omit considerations of intermittency, they will not be elaborated upon in the ensuing discussion.

\begin{figure*}
     \centering
     \includegraphics[width=1\textwidth]{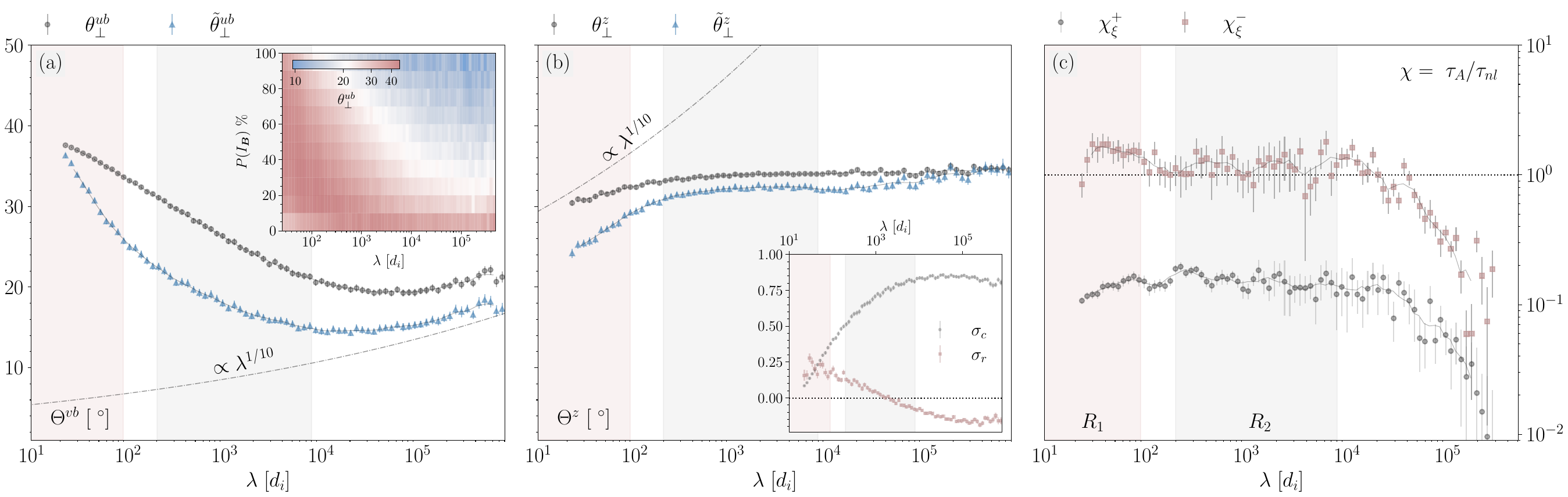}
     \caption{Alignment angles (a) $\Theta^{ub}(\lambda)$ and (b) $\Theta^{z}(\lambda)$, constrained to the range of $0^{\circ} \leq \Theta^{ub(z)} \leq 90^{\circ}$. The black curves depict alignment angles calculated with Equation \ref{eq:9}, whereas the blue curves are based on Equation \ref{eq:10}. The inset in panel (a) highlights the inverse relationship between alignment angle and field gradients by color-coding $\theta^{ub}$ across scale-dependent percentile bins of $I_{\boldsymbol{B}}$. The $i$-th bin at scale $\lambda$ is defined as $[10(i-1), 10i)$, for $i = 1,...,10$.  The inset in panel (b) displays  $\sigma_{r}(\lambda)$, in red, and $\sigma_{c}(\lambda)$, in gray. Panel (c) shows the nonlinearity parameter, $\chi^{\pm}(\lambda)$,  for outwardly ($\delta z^{+}$) and inwardly ($\delta z^{-}$) propagating waves, depicted with gray and red lines respectively. For all cases, results were derived by first calculating values for each interval independently and then computing a scale-dependent median.}
     \label{fig:align_sdda}
\end{figure*}

\par From an observational standpoint, the extent to which imbalance impacts higher-order statistics in MHD turbulence, is largely unexplored territory. Previous investigations have predominantly focused on categorizing findings based on wind speed or analyzing mixed Fast/Slow streams \citep{Horbury_1997, Mengeney_2001, Salem_intermittency, chhiber_subproton, Wu_2023}, with a recent shift in interest towards magnetic compressibility \citep{2022_Palacios}. However, these studies often do not explicitly detail the degree of Els\"asser imbalance in the dataset, making direct comparisons with our observations challenging. Nonetheless, it is worth pointing out that the scaling profiles observed in the $R_2$ range, especially for the parallel and perpendicular components, qualitatively align with results presented in \cite{osman_intermittency_2012}.

\par In terms of numerical simulations, both homogeneous and inhomogeneous setups have been employed to examine the influence of imbalance on MHD turbulence statistics. Studies have primarily concentrated on spectral properties, revealing inertial range scalings of $\alpha_{\lambda} = -3/2$ \citep{Perez_2012, Chandran_Perez_2019, grappin2022modeling, meyrand2023reflectiondriven}. 

\par \cite{2023_Chen_compres} investigated the effects of Els\"asser imbalance on higher-order statistics using both homogeneous and EBM simulations, each initialized with varying degrees of imbalance. It was shown that EBM simulations initialized with $|\sigma_c| \approx 1$ result in higher-order moment scaling exponents consistent with the predictions of the \citetalias{chandran_intermittency_2015} model. Conversely, simulations with lower values of $\sigma_c$ exhibited scaling exponents showing a linear dependence on order $n$. The study also highlighted significant differences in higher-order scaling exponents between homogeneous and EBM simulations, even when initialized with identical levels of imbalance. For example, scaling exponents in homogeneous runs with $\sigma_c \approx 0$ were found to be convex functions of order and closely resembled the \citetalias{chandran_intermittency_2015} model, in line with previous investigations \citep{Chandran_2015, 2015Mallet, 2022_Palacios}. Additionally, \cite{2023_Chen_compres} noted that in EBM simulations, the scaling properties displayed variations when higher-order moments were calculated from increments sampled in directions other than radial.

\par The latter observation is consistent with the findings of \cite{2015_verdini}, who conducted a comparative analysis of balanced homogeneous and EBM simulations. They observed that while the homogeneous simulations displayed three-dimensional anisotropy, in agreement with \citetalias{boldyrev_2006}—a finding further corroborated by \cite{2016_mallet}—the EBM simulations demonstrated axisymmetry relative to $\boldsymbol{B}_{\ell}$ and did not exhibit three distinct inertial range scaling laws. Specifically, EBM simulations with increments measured along the radial direction demonstrated spectral scalings of $\alpha_{\lambda} \approx \alpha_{\xi} \approx -3/2$ in both the perpendicular and displacement components. However, measurements in the transverse direction revealed scalings of $\alpha_{\lambda} \approx \alpha_{\xi} \approx -5/3$. In both cases, the parallel component lacked convincing scaling properties, although it exhibited a slightly steeper spectrum compared to the perpendicular components.

\par These findings lead to the following interpretation: The large-scale flow, being radial rather than uniform, cannot be negated by a Galilean transformation. Consequently, the expansion introduces an additional axis of symmetry and fosters a scale-dependent competition between the mean-field and radial axes \citep{1973_Volk}.  Intuitively the effects of the expansion should be important when the non-linear time, $\tau_{nl}$ is slower than the expansion of the solar wind, $\tau_{exp} = R/V_{SW}$, $\tau_{exp} \leq \tau_{nl}$. Given that $\tau_{nl}$ is scale-dependent, it logically follows that the effects of expansion are more pronounced at larger scales\footnote{Note, however, that observational evidence suggests that expansion can contaminate the turbulence statistics even within inertial scales \citep{Verdini_3D_2018, Verdini_2019}}. From this discussion, it becomes evident that the expansion has the potential to modify the local 3D anisotropy in a scale-dependent manner.

At the resolution currently achievable in (R)MHD simulations, meaningful comparisons are possible with the larger-scale end of the inertial range, $R_2$, spanning $200-6000d_i$. Focusing on $\zeta_{\lambda}$ and setting aside anisotropy, our findings are consistent with \cite{2023_Chen_compres}'s imbalanced EBM results and, consequently, with the models of \citetalias{chandran_intermittency_2015} and \citetalias{Mallet_2017}. Nevertheless, the observed discrepancies in the scaling exponents of the parallel and displacement components within $R_2$ could hint at the influence of expansion effects, suggesting a scenario where the dominant axis of symmetry is a mix of both $B_{\ell}$ and the radial axis, with the contribution of each being scale-dependent.

\par Shifting focus to the scaling exponents in $R_1$ (spanning $8-100d_i$), recent theoretical work suggests that under conditions of strong imbalance, generalized helicity conservation may hinder turbulent energy transfer to kinetic scales \citep{PassotSulemTassiPOP2018,PassotSulemJPP2019,meyrand_2021,PassotSulemLavederJPP2022}. The ``helicity barrier'' effect could influence both spectral and potentially higher-order moment scalings at the smaller end of the inertial range. Homogeneous hybrid-kinetic simulations, initialized under strong imbalance conditions to capture this effect \citep{2022_Squire_helicity}, exhibit spectral exponents for the parallel and perpendicular magnetic field components, $a_{\ell_{||}} \approx -2$ and $a_{\lambda} \approx -5/3$ respectively, in line with observations in the $R_1$ range.

\par While our current data does not definitively link our observations to the helicity barrier effect, it underscores the necessity for more comprehensive numerical studies focusing on the higher-order statistics of strongly imbalanced turbulence.

\par In conclusion, this discussion underscores the significant impact of imbalance and expansion effects on MHD turbulence statistics, indicating that the solar wind might not provide an ideal laboratory for evaluating homogeneous MHD turbulence models. This realization calls for a prudent application of homogeneous, balanced turbulence models in the analysis of solar wind observations and highlights the imperative for more sophisticated theoretical modeling and refined data interpretation techniques.

\subsection{ Critical Balance (CB)}\label{subsec:CB}

\par Using balanced RMHD simulations, \cite{2015Mallet} demonstrated that although the distributions of $\tau_{nl}$ and $\tau_{A}$ are not self-similar, their ratio, $\chi$, maintains a scale-invariant distribution within the inertial range. \cite{Chhiber_2020_timescales} employed balanced incompressible MHD simulations to show that while the $\chi$ distribution peaks at $\chi \approx 1$, it is asymmetric and skewed towards $\chi \geq 1$. Further reinforcing the \cite{Chhiber_2020_timescales} findings, \cite{Oughton_2020_cb} highlighted that, despite RMHD simulations producing results claimed to support CB, the similarity between RMHD’s $\chi \geq 1$ requirement and the CB condition of $\chi \approx 1$ has led to some confusion in differentiating these two theoretical frameworks.

\par \cite{chen_2016} utilized an extensive dataset of fast wind streams with moderate cross-helicity ($\sigma_c \approx 0.6$) from the outer heliosphere to investigate the scale dependence of the non-linearity parameter. They found $\chi$ to be scale-independent across the inertial range, maintaining a value around $\chi \sim 1$. Due to the lower resolution in velocity data and a less pronounced imbalance compared to our dataset, they assumed identical statistical properties for the two Elss\"asser fields, enabling them to estimate $\chi = (\ell_{||} / \lambda )(\delta b/V_{A})$ using solely magnetic field data.

\par However, numerical simulations by \cite{Berenyak_2008, Beresnyak_2009B} suggest that with increasing imbalance, statistical properties (i.e., amplitudes, coherence lengths) of the two Els\"asser species diverge progressively. Considering the strong imbalance in our dataset, we employed a more refined approach, computing $\chi^{\pm} = (\ell^{\pm}_{||, \lambda}/ \lambda^{\pm})/(\delta z_{\lambda}^{\mp}/V_{a})sin \theta^{z}$.

Our analysis reveals a strong cascade for the inwardly propagating waves, with $\chi^{-}$ remaining scale-independent across the inertial range, maintaining a value slightly above unity. In contrast, the outwardly propagating waves exhibit a weaker cascade, with $\chi^{+}$ increasing from $0.1$ at $\lambda \approx 10^{4} d_i$ to $0.2$ at $\lambda \approx 10^{2} d_i$

At this point, it's important to recognize two key factors that might affect the accuracy of our $\chi^{\pm}$ estimates. First, there's a prevailing assumption that $z^{\pm}$ structures are primarily sheared by counter-propagating $z^{\mp}$ wavepackets of similar perpendicular scale, which implies a cascade that is local in $\lambda$. However, this notion is challenged by the work of \cite{Schekochihin_2022} who put forward a model of imbalanced turbulence that consists of two strong, ``semi-local'' cascades: one local in $\lambda$ for the stronger field and another, local in $\ell_{||},$ for the weaker field. The implications of this model cast doubt on our estimates of $\chi^{\pm}$ that is predicated on the concept of scale locality.

\par Furthermore, the transition from weak to strong turbulence, characterized by a change from $\chi \ll 1$ and $\alpha_{\lambda} \approx -2$ to $\chi \approx 1$ and $\alpha_{\lambda} \approx -3/2$, is a pivotal aspect of the CB discussion. This transition, observed in balanced RMHD shell-model simulations \cite{Verdini_transition}, 3D incompressible MHD simulations \citep{Meyrand_transition}, and recently in the Earth's magnetosheath \citep{Zhao_2023_transition}, remains unreported in the solar wind \footnote{Even though such a transition was speculated in recent works \citep{telloni_transition, Wu2022OnTS, Sioulas_2023_anisotropic}}. 

Our results indicate a transition in the cascade of the ingoing wave from weak to strong turbulence, with $\chi^{-} > 1$ at $\lambda \approx 3 \times 10^{4} d_i$, yet without capturing the anticipated scaling transition. We speculate, below, that this could be related to the effects of ``anomalous coherence''. More specifically, in the context of homogeneous MHD, nonlinear interactions between counterpropagating waves are uncorrelated and transient, limited to the duration of encounters. In the solar wind, however, nonlinear dynamics are complicated by \textit{anomalous coherence}, a phenomenon arising from non-WKB reflection of outwardly propagating fluctuations \citep{velli_turbulent_1989, Velli_1990, Hollweg_Isenberg_2007}. A key aspect of this effect is the presence of an ``anomalous'' reflected component, $z^{-}_{a}$, in addition to the ``classical'' component $z^{-}_{c}$, which remains stationary relative to the $z^{+}$ frame, coherently shearing it throughout its lifetime \citep{Verdini_2009ApJ, Chandran_Perez_2019}. Under strong imbalance and inhomogeneity, $z^{-}_{a}$ can assume a leading role at large scales, altering the phenomenology of the energy cascade and leading to distinct spectral behaviors: a $1/f$ scaling for outwardly propagating waves and an $f^{-3/2}$ scaling for inwardly propagating waves \citep{velli_turbulent_1989, Perez_Chandran_2013, meyrand2023reflectiondriven}.

\par In summary, our results suggest that at inertial scales, outgoing waves experience a weak cascade, while ingoing waves undergo a strong one, closely resembling the CB condition ($\chi^{-}_{\xi} \approx 1$). However, given the complexities previously discussed and the uncertainties inherent in our measurements, we advise interpreting these findings with a degree of caution.

\subsection{ Scale Dependent Dynamic Alignment (SDDA)}\label{subsec:SDDA}

\par Several numerical investigations into homogeneous (R)MHD \citep{masson_2006, Perez_2012, Perez_2014, 2022_cerri} have provided ample evidence for alignment signatures spanning a significant portion of the inertial range. However, \cite{2012_Beresnyak, 2014_Beresnyak} have suggested that this observed alignment increase may be a finite-range phenomenon closely linked to dynamics at the outer scale.

\par Observational studies using data sampled at 1 AU have provided evidence of alignment at large, energy-containing scales. However, it has been observed that this trend towards increasing alignment diminishes at inertial scales \citep{Podesta_2009, Hnat_scale_free_2011, wicks_alignmene_2013, Parashar_2019}. This trend persists even in data intervals specifically chosen to mitigate the effects of solar wind expansion \citep{Verdini_3D_2018}. It has been noted, however, that small errors in velocity vector measurements, due to instrumental limitations, can lead to significant errors in alignment angle measurements, even at large scales \citep{PODESTA_2009_SDDA}.

\par Recently, \cite{Parashar_2020} explored the scale dependence of several alfv\'enicity diagnostics during E1 of PSP. Their findings suggest that $\sigma_c$ starts decreasing, with $\sigma_r$ increasing, at scales considerably larger than those observed at 1 AU \citep{Podesta_2010}, despite high alfv\'enicity at large scales. This observation aligns with HELIOS observations \citep{1990_TU} and has been attributed to the substantial energy found in velocity shears in the inner heliosphere \citep{Ruffolo2020}. Specifically, shear disrupts an initial spectrum of high cross helicity by injecting equal amounts of the two Els\"asser energies \citep{Roberts_1987, Goldstein_1989, roberts_velocity_1992}.

\par The scale-dependence of the alignment angles correlates directly with that of $\sigma_c$ and $\sigma_r$, as only two out of these four quantities are independent. Specifically, the formal relationship between imbalance, residual energy, and alignment, as described by $cos\theta^{z}_{\perp} = \sigma_r/ (1- \sigma_c^{2})^{1/2}$ and $cos\theta^{ub}_{\perp} = \sigma_c/ (1- \sigma_r^{2})^{1/2}$ , indicates that the development of both Els\"asser imbalance and residual energy, i.e., a monotonic increase in $|\sigma_c|$ and $|\sigma_r|$ towards smaller scales, is necessary for SDDA to emerge towards smaller scales \citep{wicks_alignmene_2013, Schekochihin_2022}.

\par Such trend is evident in Figure \ref{fig:align_sdda}, where alignment signatures become apparent only when $\sigma_c$ exhibits a monotonic increase at scales $\lambda < 10^{4} d_i$, or when $\sigma_r$ becomes positive, leading to a monotonic increase in $|\sigma_r|$ at scales $\lambda < 2\times 10^{3} d_i$. Although instrumental noise might influence the latter trend, as discussed in \cite{bourouaine_turbulence_2020}, the observed behavior at energy injection scales aligns with 1 AU observations \citep{wicks_alignmene_2013}. In contrast, in the inertial range, $\theta^{z}(\approx 35^{\circ})$ remains roughly scale-independent.

\par These observations raise a critical question: Is the observed scale-dependence of the alignment a reflection of actual physical processes, or might it simply be a consequence of instrument characteristics?

 \par Potential physical mechanisms may encompass interactions between compressive and non-compressive modes \citep{Cho_Lazarian_2003, Chandran_2005}, ideal MHD instabilities manifesting in the solar wind, including the Kelvin-Helmholtz instability \citep{Malagoli_1996_KH}, the cessation of the aligning cascade due to the tearing instability \citep{Mallet_2017, Boldyrev_Loureiro_2017, Comisso_2018_alignment}, or even the solar wind's inherent inhomogeneity resulting in non-WKB reflections and a reduction in cross-helicity.

\par The extent to which instrumental noise influences these observations remains a crucial, yet unresolved, concern, emphasizing the necessity for a careful interpretation of observational data. While our analysis cannot definitively assert the nature of SDDA at small scales, it provides compelling observational evidence suggesting an inverse relationship between alignment angles and the intensity of field gradients, thereby corroborating the numerical results of \cite{2015Mallet}.  As such, our observations lend support to the \citetalias{chandran_intermittency_2015} model, suggesting that the physical basis of alignment lies in the mutual shearing of Els\"asser fields during imbalanced collisions between counterpropagating wave packets.

Before concluding, another topic related to our findings deserves further discussion. Our observations indicate strong anti-alignment between Els\"asser species at larger scales, with $\sigma_c \approx 1$ and typically small and negative $\sigma_r$, implying, as suggested by \cite{wicks_alignmene_2013}, $cos\theta^{z}_{\perp}\approx 180^{\circ}$ at $\lambda \approx 10^{4} d_i$. It is natural, then, to inquire how this picture would be modified in the case of globally balanced streams. For instance, as emphasized in \citetalias{chandran_intermittency_2015}, outer-scale fluctuations in the context of globally balanced turbulence are not expected to be strongly aligned. This suggests that while imbalanced turbulence may exhibit alignment saturation at larger scales, balanced turbulence has the potential for further alignment, assuming that $\sigma_c$ has room to increase at smaller scales. Thus, the dynamics of alignment across scales in balanced turbulence present an intriguing area for future research, particularly in comparing the extent of this alignment process with that in imbalanced streams. This aspect warrants further investigation and will be the focus of future work.

\subsection{Exploring the Efficacy of the \texorpdfstring{\citetalias{chandran_intermittency_2015} \& \citetalias{Mallet_2017}}{Chandran et al. 2015 \& Mallet et al. 2017} Models in the Context of Imbalanced Turbulence}\label{subsec:CSM15_model}

Our results indicate that the intermittent scalings of the $n$th-order conditional structure functions in the direction perpendicular to both the local mean field and the fluctuation directions closely align with the theoretical frameworks proposed by \citetalias{chandran_intermittency_2015} and \citetalias{Mallet_2017}. However, our dataset exhibits notable characteristics that diverge from the assumptions underlying these models. These include strong imbalance in the fluxes of the Els\"asser species and a prevalence of eddies conforming to a field-aligned tube topology, which, on average, do not display increasing alignment towards smaller scales. At first glance, these observations might seem contradictory to the expectations set forth in the aforementioned models. Nevertheless, upon closer examination of the model's fundamental assumptions, it becomes apparent that these empirical findings are not inconsistent with the model’s theoretical framework.

In Section \ref{sec:background}, we examine the foundational assumptions of the \citetalias{chandran_intermittency_2015} and \citetalias{Mallet_2017} models. These assumptions encompass (1) negligible cross helicity at energy injection scales and (2) the formation of eddies following a current sheet topology, characterized by a volume filling factor $f_{cs} \propto \lambda$, alongside the dynamic alignment of vector field fluctuations at smaller scales. The models incorporate alignment as an intermittency effect, resulting in (1) the inherent introduction of local Els\"asser imbalance and (2) the lack of a requirement for the ``average''  eddy to adhere to a 3D anisotropic current sheet topology. The models propose that it is the eddies in the tails of the PDFs that are expected to exhibit this topology, showing increased alignment at smaller scales. In essence, an inverse correlation between alignment and the intensity of field gradients is anticipated at any given scale.

When second-order moments are employed to examine the statistical shape of eddies, the high-amplitude, current-sheet-like structures found in the tails of the increment PDFs are typically obscured during the averaging process by more common, lower-amplitude, tube-like eddies. However, as higher moment orders are considered, these high-amplitude structures gain more prominence in the ensemble average, thereby significantly impacting the profile of the scaling exponents.

Given that fluctuations in the perpendicular component are energetically dominant, any ``side effects'' stemming from the disruption of anisotropy due to expansion, and possibly from constraints imposed by spherical polarization \citep{Matteini_2024}, do not seem to significantly impact the scaling characteristics of the perpendicular component. However, they do appear to influence the scaling profiles of the displacement and parallel components. Hence, incorporating these effects into the existing homogeneous models could potentially enhance the scaling predictions of these models and offer valuable insights into the nature of intermittent turbulence in the expanding solar wind.

\subsection{Observations of the outer scale}\label{subsec:Observations}

\par At large energy-containing scales, as depicted in Figure \ref{fig:wavevector_anisotropy}, the parallel and perpendicular components display rough equipartition in fluctuating energy. However, the displacement component appears somewhat more energetic in comparison. Consequently, eddies deviate from isotropy, exhibiting a subtle compression along the fluctuation direction.

Our findings contrast with those of \cite{Chen_2012ApJ}, who observed eddy elongation along the displacement direction ($\hat{\xi}$) at large scales in fast solar wind data at 1.4 AU.

To delve deeper into these findings,  \cite{2015_verdini} conducted a comparative analysis using 3D MHD homogeneous and EBM simulations. They found isotropic eddies at energy injection scales in non-expanding simulations. However, in expanding simulations where increments were measured along the radial direction (typical for single-spacecraft missions at 1 AU), results aligned with those reported by \cite{Chen_2012ApJ}. Interestingly, large-scale anisotropy disappeared, and eddies appeared isotropic when increments were measured in non-radial directions. This was interpreted as an effect of expansion, which preferentially dampens the radial component of magnetic field fluctuations relative to the azimuthal ones, confining fluctuations primarily to the plane orthogonal to the radial direction and leading to an anisotropic energy distribution among the field components. This phenomenon was observed in 3D EBM simulations \citep[see e.g., ][]{Dong_2014}.

In a more recent study, \cite{Verdini_3D_2018} analyzed a decade of data from the Wind spacecraft and identified a correlation between large-scale anisotropy and variance anisotropy, defined as $E = (b^{2}_{T} + b^{2}_{N})/b^{2}_{R}$, where $b$ represents the rms amplitude of fluctuations. They found that intervals corresponding to the ``strong'' expansion dataset (defined by $E>2$) exhibited eddy elongation along the displacement direction, consistent with \cite{Chen_2012ApJ}. Conversely, intervals from the ``weak'' expansion dataset (with $E\leq2$) showed eddy elongation along the perpendicular direction ($\hat{\lambda}$).

Based on these results, \cite{Verdini_3D_2018} suggested that PSP, due to its unique orbit-- allowing measurements perpendicular to the radial direction during its near-sun phase-- would detect isotropic eddies at energy injection scales.

To overcome the limitations of single spacecraft measurements, \cite{Vech_2016} adopted a multispacecraft approach, enabling the separation of measurements along both radial and transverse directions. This strategy facilitated the isolation of expansion, solenoidality, and the mean magnetic field effects. Their investigation underscored the dominant role of the solenoidality constraint \citep{Turner_2011} over expansion in contributing to the observed variance anisotropy, E \citep[see e.g.,][]{Horbury_2001}. They further noted that while some anisotropy, as observed with radial increments, stemmed from expansion, a reduced yet noticeable degree of anisotropy persisted when measurements were conducted along the transverse direction. This suggests the involvement of additional mechanisms in shaping the large-scale 3D anisotropy.

\par Recent in-situ observations have shed light on the decay observed in the radial component within the inner heliosphere, indicating that it cannot be solely attributed to expansion effects. Specifically, analysis of data from PSP and HELIOS by \cite{Tenerani_2021} demonstrated that the rms of fluctuations in the radial component decays at a slower rate compared to that of the perpendicular component. This phenomenon was further clarified by \cite{Matteini_2024}, who attributed it to the tendency of magnetic field fluctuations in the solar wind to evolve towards a state of spherical polarization. The spherical polarization imposes constraints on the radial component's rms fluctuations, leading to a decay described, particularly at large scales, by $b_{r} \sim \boldsymbol{b}/2B$ \citep[see also,][]{Squire_2020, Mallet_2021}.

Returning to our findings, we observe only a slight deviation from isotropy at the outer scale, which is notable considering that our dataset falls within the strong expansion category, with an average variance anisotropy of $E \approx 4.26 \pm 3.51$. Additionally, the sampling angle consistently falls within the range of $\theta_{R} \in [160^{\circ}, 180^{\circ}]$, indicating quasi-radial sampling. Despite identifying several intervals among multiple PSP encounters that exhibit isotropic large-scale eddies, no clear correlation has emerged from this preliminary analysis between prevailing plasma signatures (e.g., $E$, $\sigma_c$, $\sigma_r$, $\theta_{R}$) and the occurrence of such intervals.

\par Taking into account the subtleties revealed by the recent observations discussed above, we must recognize our current inability to offer a satisfactory explanation for the observed configuration of the large-scale eddies in our dataset. However, the differences noted in the near-Earth and near-Sun eddies could indicate preferential damping in the fluctuations of the displacement component of the magnetic field. To clarify this aspect, it would be worthwhile to explore the development of the large-scale eddies at varying heliocentric distances.

\subsection{Isotropization of eddies at small scales}\label{subsec:eddie_isotrop}

Figure \ref{fig:3D_eddies} illustrates that within the $R_1$ range, the eddies display increasing anisotropy, resembling ribbon-like structures towards smaller scales. However, the trend of increasing aspect ratio ceases at $\lambda \approx 2 d_i$, at which point the eddies transition toward a quasi-isotropic state. In the following, we discuss two potential mechanisms that could explain these observations.

\par The observed transition towards isotropy at smaller scales is consistent with the idea that thin, long-lived current sheets generated by the turbulent cascade can be disrupted by the tearing instability and subsequent reconnection \citep{Furth_tearing}. Specifically, when the maximum growth rate of the \cite{Coppi_1976} mode, $\gamma_{t}$, representing the fastest tearing mode in an MHD sheet, becomes comparable to the non-linear cascade time $\tau_{nl}$ ($\gamma_{t} \tau_{nl} \gtrsim 1$), the stability of the current sheets is compromised \citep{Pucci_Velli_2014, Uzdensky_Loureiro_2016}. The instability leads to the fragmentation of the dynamically forming sheets into flux ropes, which exhibit isotropy in the plane perpendicular to the magnetic field. This phenomenon is expected to occur at scale, $\lambda_{D}$, beyond which the nature of the MHD cascade undergoes a significant transformation. The disruption of the current sheets affects the dynamic alignment, accelerating the turbulent cascade and resulting in a noticeable steepening of the power spectrum \citep{2017_Mallet_tearing, 2017_Loureiro, Boldyrev_Loureiro_2017}.

\par While this mechanism appears feasible for balanced turbulence, as evidenced by observations in the solar wind \citep{Vech_mallet_2018} and more recent findings in 3D fully-compressible \citep{Dong_22_largest_mhd_turb} and reduced \citep{2022_cerri} MHD simulations, it's important to acknowledge that the cutoff of the inertial range in imbalanced turbulence might also be influenced by kinetic effects. Therefore, despite the observation of a sub-ion-scale range mediated by magnetic reconnection in 2.5D hybrid-kinetic simulations \citep{CerriCalifanoNJP2017,FranciAPJL2017}, other kinetic-scale mechanisms in imbalanced turbulence could potentially contribute to eddy isotropization. For instance, an alternative explanation for the isotropization of the eddies at small scales could be associated with the recently discovered ``Helicity Barrier'' mechanism \citep{PassotSulemTassiPOP2018,PassotSulemJPP2019,meyrand_2021,PassotSulemLavederJPP2022}.

As discussed in the introduction, strongly magnetized (low-beta) collisionless plasmas exhibit nonlinear conservation of both energy and cross helicity. However, the conserved quantity in reality is termed ``generalized helicity.'' At $k_{\perp} \rho_i \lesssim 1$, this corresponds to the cross helicity following a forward cascade, which conservatively transforms into magnetic helicity at $k_{\perp} \rho_i \gtrsim 1$, undergoing an inverse cascade. Consequently, an imbalanced cascade arriving from the inertial range faces a complication—the sudden need to reverse the direction of the generalized helicity cascade. The helicity barrier impedes the viability of a constant-flux cascade, leading to an accumulation of energy in the stronger Els\"asser field. This accumulation shortens $\tau_{nl}$, reducing the parallel correlation length, in line with the CB theory, to the extent that turbulent energy is redirected into an (ICW; \citep[see, ][]{Stix_1992})  spectrum. This mechanism opens up a new dissipation channel via the ion-cyclotron resonance \citep{Squire_2022}.

ICWs are commonly observed in the nascent solar wind, particularly during intervals marked by (anti)alignment between the mean magnetic field and solar wind flow direction \citep{Bowen_2020c}. A strong correlation exists between the presence of ion-scale waves and the level of imbalance of fluctuations at inertial scales \citep{Zhao_2022, bowen2023mediation}. The presence of ICWs can significantly impact the power spectra of magnetic fields at ion kinetic scales \citep{Bowen_2020c, Shankarappa_2023}. Specifically, the bump observed in the parallel spectrum just before the transition region has been attributed to the presence of ICWs, suggesting that the isotropization of the eddies could be a consequence of the helicity barrier mechanism.

The main emphasis of this analysis is on the inertial and energy injection scales. The extended interval size allows for more reliable estimates of second-order moments and, consequently, the anisotropic curves presented in Figure \ref{fig:wavevector_anisotropy}. However,  no effort has been made to account for the energy contribution ICWs. Due to the strong correlation between the presence of ion-scale waves and Els\"asser imbalance, simply disregarding intervals with ICW wave signatures \citep{Duan_2021, Zhang_2022} may impede our investigation into kinetic-scale turbulence statistics in strongly imbalanced intervals. An alternative approach would entail identifying and eliminating the energy attributed to ICWs from the observed energy spectrum \citep[see e.g.,][]{Shankarappa_2023, Wang_2023_anisotropy}, and examining the resulting anisotropy based on the parallel and perpendicular spectra. This will be the focus of an upcoming study.

 \subsection{Can the trace PSD be interpreted as the perpendicular PSD ?}\label{subsec:Trace_perpendicular}

The theoretical models discussed in Section \ref{sec:intro} provide scaling predictions for the parallel, perpendicular, and displacement components of fluctuating fields.

However, the angle between the solar wind flow and the magnetic field, as observed by spacecraft, can significantly influence whether fluctuations in measured quantities vary parallel or perpendicular to the magnetic field. Due to the Parker Spiral configuration \citep{Parker58}, the baseline value of the angle between the solar wind flow and the magnetic field, denoted as $\Theta_{VB}$, increases with heliocentric distance. Consequently, spacecraft measuring magnetic field fluctuations at 1 AU are more likely to detect fluctuations perpendicular to the  mean magnetic field direction.

Considering the strongly anisotropic nature of the turbulent cascade, with the majority of power associated with perpendicular wavenumbers \citep{shebalin_matthaeus_montgomery_1983, Montgomery_1981, horbury_anisotropic_2008}, observational studies have traditionally estimated the trace PSD and interpreted these measurements as representative of the perpendicular spectrum. However, as the PSP moves closer to the Sun, both the flow and the magnetic field become predominantly radial. Consequently, PSP often detects variations parallel -resulting in a deficit of measurements perpendicular- to the magnetic field. This can impact the statistical signatures of MHD turbulence, including intermittency \citep{Sioulas_2022_intermittency}, estimates of correlation lengths \citep{Cuesta_anisotropy}, etc. As shown in Figure \ref{fig:overview}, the deficit caused by the sample size of the perpendicular fluctuations due to quasi-parallel sampling can also result in strong deviations between the perpendicular and trace PSD. This effect becomes even more important at later PSP encounters, and thus caution should be exercised when trying to utilize the trace PSD to compare with perpendicular PSD predictions of theoretical models.

\subsection{Assessing 2-Point Structure Functions for Small-Scale Turbulence Analysis}\label{subsec:reconnection}

 \par The departure from monofractal statistics, exemplified by the adoption of $SF^{n}_{5}$, underscores the inadequacy of the $SF^{n}_{2}$ method for statistical analysis of MHD turbulence at smaller spatial scales. Specifically, the $SF^{n}_{2}$ method lacks accuracy in capturing the scaling behavior at these scales, where steep scaling in the power spectra and higher-order moments are frequently seen. This deficiency can result in imprecise estimations of all associated intermittency metrics, leading to potentially erroneous interpretations of the nature of the turbulent cascade \citep[see also 3D kinetic simulations by ][]{2019_Cerri}.

\par The prolonged duration of the intervals examined in this study, which may not be optimal for a concentrated analysis at kinetic scales, in conjunction with the limitations posed by the presence of ICWs as discussed in Section \ref{subsec:anisotropy}, underscores the need for further investigation. Future studies should consider employing either 5-point or wavelet-derived structure functions, with a specific emphasis on distinguishing between balanced and imbalanced turbulence streams \citep[see, e.g.,][]{bowen2023mediation}. This comprehensive exploration is essential for achieving a deeper and more precise comprehension of the fractal properties of MHD turbulence at kinetic scales.

\subsection{Limitations }\label{subsec:limitations}

\par In addition to the limitations associated with velocity measurements discussed earlier, it is essential to acknowledge further inherent limitations in the analysis presented in this study.

\subsubsection{Finite sample size effects}\label{subsubsec:sample_size}

The proper study of MHD turbulence hinges on the ability to sample plasma from a common solar source, typically a single solar wind stream, and gather a sufficiently large sample size for statistical analysis. Spectral properties alone are insufficient for assessing scale invariance and fractal properties; higher-order moments are necessary. While evaluating structure functions is generally straightforward, estimating scaling exponents presents pitfalls. The primary concern arises from increased sensitivity to rare and large events as the order, p, increases. This can lead to finite sample effects dominating the analysis, especially as emphasis shifts to the poorly sampled tails of the distribution with higher orders. Consequently, higher-order moments become susceptible to outliers, rendering estimates of scaling exponents increasingly unreliable \citep{Dudok_de_wit_Samples_Rule, 2022_Palacios}. As a rule of thumb, it is generally deemed safe to compute structure functions up to a certain order, typically defined as $p_{\text{max}} = \log N - 1$, where N represents the sample size \citep{Dudok_de_wit_Samples_Rule}.

In our analysis, these challenges are further compounded by two factors. Firstly, we employ conditional analysis to estimate higher-order moments in three physically motivated directions, resulting in the exclusion of a significant portion of increments falling outside specified angle ranges. Secondly, we utilize 5-point structure functions, where the way increments are taken leads to a larger portion of the time series being discarded due to edge effects.

Moreover, PSP data introduces added complexity compared to 1AU measurements, as the distance from the Sun rapidly changes, causing large variations in the rms of fluctuations between intervals sampled at different heliocentric distances. Consequently, very long intervals cannot be utilized effectively.

Despite these limitations, the present analysis could be significantly improved by adopting the method described in \cite{2022_Palacios}, where a large sample of increments from non-contiguous solar wind streams with similar characteristics can be utilized to construct the PDFs needed to obtain higher-order moments. This approach will be the focus of future work.

\subsubsection{Scaling exponents of Els\"asser fields.}\label{subsubsec:sc_epx_zpm}

In this study, our primary focus has been on the higher-order moments derived from the magnetic field timeseries. However, it's crucial to recognize that the fundamental variables in MHD are the Els\"asser fields—rather than $\boldsymbol{B}$ and $\boldsymbol{V}$—due to their conservatively cascading energies. Indeed, the scaling predictions provided by the \citetalias{chandran_intermittency_2015} and \citetalias{Mallet_2017} models pertain to the scaling exponents of the  Els\"asser field increment moments. Therefore, a more direct comparison with these models would entail estimating the moments of increments in $\boldsymbol{z}^{\pm}$  \citep[see e.g.,][]{2022_Palacios}. However, adopting this approach would require downsampling the magnetic field timeseries to synchronize with the cadence of the velocity field data. This would lead to a notable reduction in sample size, and render the estimation of anisotropic higher-order moments unfeasible with our current dataset.

Nevertheless, as discussed in \citetalias{chandran_intermittency_2015}, the regions contributing dominantly to both types of structure functions are those where $\delta \boldsymbol{z}^{\pm}$ exhibits exceptional magnitudes. In these regions, one Els\"asser fluctuation, e.g., $\delta \boldsymbol{z}^{+}$, typically dominates over the other, leading to $\delta \boldsymbol{b} \approx  (1/2) \delta \boldsymbol{z}^{+}$. Therefore, given the significant imbalance in our dataset, the scaling exponents estimated for the magnetic field timeseries can provide a reasonable approximation for the scaling exponents of the dominant (outgoing) Els\"asser field.

\subsubsection{Switchbacks}\label{subsubsec:SBs_affect?}

The near-Sun solar wind environment is characterized by the prevalent occurrence of Switchbacks, a subset of predominantly Alfv\'enic fluctuations with amplitudes significant enough to cause the magnetic field to reverse its direction abruptly, resulting in a local field polarity reversal and a corresponding radial velocity jet \citep{Matteini_2014, 2018_horbury_switchbacks, bale_highly_2019}.

The question arises as to what extent these sudden reversals impact our ability to accurately estimate the local magnetic field, and consequently, the scaling exponents of the parallel and displacement components.

While excluding switchbacks from the analysis could potentially address this concern, it's noteworthy that the majority of our samples for the two perpendicular components originate from substantial kinks in the magnetic field time series, as these events lead to large $\Theta_{BV}$ angles. Therefore, no attempt has been made to further clarify this aspect. However, it is reassuring to note that the scaling exponent profiles obtained for a substantial dataset of imbalanced Wind observations at 1 AU, where significant kinks in the magnetic field time series typically diminish and  switchbacks/switchback patches transition into microstreams \citep{Horbury_2023, soni2024switchback}, are qualitatively consistent with those reported in the current analysis.

\section{Conclusions and Summary}\label{sec:Conclusions}

We analyzed in-situ observations from a highly Alfv\'enic stream captured during Parker Solar Probe's first perihelion to assess the predictions of MHD turbulence models grounded on the principles of ``Critical Balance'' and ``Scale-Dependent Dynamic Alignment''. Our objective was to assess the extent to which the conjectures made and predictions derived by these models align with in-situ solar wind observations and establish solid observational benchmarks for the testing and refinement of MHD turbulence phenomenologies.

 \vspace{1em}
The main findings of our study can be summarized as follows:

  \vspace{1em}
    At the outer scale,  $\lambda \gtrsim 2 \times 10^{4} d_i$, we find:
 
    \vspace{0.5em}
    (a1) Both (out)ingoing waves undergo a weak cascade, $\chi^{\pm} < 1$, that strengthens towards smaller scales. The trend is concurrent with tighter scale-dependent dynamic alignment (SDDA) of fluctuations, a monotonic increase in cross-helicity ($\sigma_c$), and a shift towards more negative residual energy ($\sigma_r$) values

     \vspace{0.5em}
    (a2) The ingoing waves transition to a strong cascade ($\chi^{-} \gtrsim 1$) at $\lambda \approx 3 \times 10^{4} d_i$; the associated spectral scalings deviate from the expected weak-to-strong turbulence transitions. We explore the possibility that ``anomalous coherence' effects may account for this discrepancy in Section \ref{sec:Disussion}.

\vspace{1em}

 \par  The domain canonically identified as the inertial range is comprised of two distinct sub-inertial segments that exhibit distinct turbulence statistics.

 \vspace{1em}
 For the subinertial range spanning $200-6000d_i$ and termed $R_2$ we find:

    \vspace{0.5em}

    (b1) Spectral scaling indices for components parallel to the local mean field, fluctuation (displacement), and perpendicular directions assume values of $\alpha_{\ell_{||}} = -1.66 \pm 0.05$, $\alpha_{\xi} = -1.56 \pm 0.08$, and $\alpha_{\lambda} = -1.49 \pm 0.03$, respectively.

    \vspace{0.5em}
    (b2) The ``average'' eddy assumes a field-aligned tube topology.

    \vspace{0.5em}
    (b3) The alignment angle $\Theta^{ub}$ between velocity and magnetic-field fluctuations monotonically increases towards smaller scales, while the alignment $\Theta^z$ between the Els\"asser fields remains roughly scale independent ($\approx 35^\circ$). In both cases, an inverse relationship between alignment angles and the intensity of field gradients is observed, suggesting that the physical basis of alignment lies in the mutual shearing of Els\"asser fields during imbalanced collisions between counterpropagating wave packets, as suggested in \citetalias{chandran_intermittency_2015}.

    \vspace{0.5em}
    (b4) The cascade is strong for inwardly propagating waves ($\chi^{-}\gtrsim 1$) but weak for outwardly propagating ones, with $\chi^{+}$ increasing from $0.1$ to $0.2$ as scales decrease from $\lambda \approx 10^{4} d_i$ to $10^{2} d_i$.

    \vspace{0.5em}
    (b5) The scaling exponents of the structure functions perpendicular to both $\boldsymbol{B}_{\ell}$ and the fluctuation direction conform to the theoretical models of \citetalias{chandran_intermittency_2015} and \citetalias{Mallet_2017}. However, the scaling profile in the parallel and displacement components deviates from theoretical predictions, possibly due to contamination from expansion effects.

    \vspace{1em}
    \par For the subinertial range spanning $10-100d_i$ (termed $R_1$), we find:

    \vspace{0.5em}
     (c1) The spectrum steepens, with spectral scaling indices for components parallel to the local mean field, fluctuation (displacement), and perpendicular directions assuming values of $\alpha_{\ell_{||}} = -1.97 \pm 0.05$, $\alpha_{\xi} = -1.94 \pm 0.06$, and $\alpha_{\lambda} = -1.64 \pm 0.04$, respectively.

     \vspace{0.5em}
     (c2)  A shift from isotropy in the plane perpendicular to $\boldsymbol{B}_{\ell}$ becomes evident, indicating a shift in eddy structures from tube-like to ribbon-like, $\ell_{||} \gg \xi \gg \lambda$. While signatures of increasing SDDA are observed, the result is potentially susceptible to errors in particle data measurements.
     
     \vspace{0.5em}
     (c3) The scaling exponents of the parallel and displacement components are a linear function of order, while the perpendicular component exhibits a weakly non-linear scaling profile. An overall transition towards ``monofractal'' statistics and a weakening of intermittency, compared to $R_2$, are evident.
    
    \vspace{1em}
    (d) At $\lambda \approx 8 d_i$, the increase in aspect ratio ceases, and the eddies transition to a quasi-isotropic state. This shift might be a signature of the tearing instability, potentially leading to reconnection of the thin current sheets, or it could result from turbulent energy being channeled into an ion-cyclotron wave spectrum, consistent with the ``helicity barrier'' effect.

    \vspace{0.5em}
     (e) The 2-point structure function method $SF^{n}_{2}$ is inadequate for capturing the essential properties of the turbulent cascade, at smaller scales. To accurately characterize steeper power laws at smaller spatial scales, the use of a more sophisticated method such as the 5-point structure function $SF^{n}_{5}$ is essential.

\vspace{1em}

\par While our study doesn't delve into the direct application of diagnostics for expansion and imbalance effects, it's interesting to note that preliminary findings using data from the Wind mission show a notable correspondence with the results from EBM simulations conducted by \cite{2023_Chen_compres}. Specifically, the scaling exponents in $R_2$ are consistent with \citetalias{chandran_intermittency_2015} when $\sigma_c \approx 1$, and become a linear function of order as the imbalance decreases to $\sigma_c \approx 0$. Furthermore, extending the work of \cite{Verdini_3D_2018} to higher order moments it is found that when intervals are selected in such a way as to minimize the expansion effects the scaling exponents in all three components are in striking agreement with those predicted by the \citetalias{Mallet_2017} model (Sioulas et al. 2024, in progress).

In summary, our findings suggest that the models proposed by \citetalias{chandran_intermittency_2015} and \citetalias{Mallet_2017}, which integrate SDDA as an intermittency effect and account for local imbalance, possess the essential elements for a successful phenomenological representation of imbalanced MHD turbulence. This assertion stems from several key observations: firstly, the models provide scaling predictions for higher-order moments in the perpendicular component of the magnetic field that align well with our in-situ observations. Secondly, an inverse relationship between alignment angles and the intensity of field gradients suggests that the alignment mechanism originates from the mutual shearing of fields during imbalanced collisions of wavepackets. However, it's worth noting that certain aspects of solar wind turbulence, such as the presence of two sub-inertial ranges and anisotropic signatures, remain unaddressed by the models. This suggests that incorporating additional effects, such as accounting for inhomogeneity or the spherical polarization of fluctuations, could enhance the models' scaling predictions.

\vspace{12pt}
\begin{acknowledgments}

N.S. acknowledges useful conversations with Michael Stevens about the use of SPC data. We acknowledge PSP/FIELDS team
(PI: Stuart D. Bale, UC Berkeley) and PSP/SWEAP team (PI: Justin Kasper, BWX Technologies) for the use of data.
This research was funded in part by the FIELDS experiment on the Parker Solar Probe spacecraft, designed and developed under NASA contract NNN06AA01C; the NASA Parker Solar Probe Observatory Scientist grant NNX15AF34G and the  HERMES DRIVE NASA Science Center grant No. 80NSSC20K0604. 
The instruments of PSP were designed and developed under NASA contract NNN06AA01C. BC acknowledges the support of NASA grant 80NSSC24K0171.

\end{acknowledgments}

\software{Python \citep{van1995python}, SciPy \citep{2020SciPy-NMeth}, Pandas \citep{mckinney2010data},  Matplotlib \citep{Hunter2007Matplotlib}, Pyspedas \citep{angelopoulos_space_2019}, \citep{MHDTurbPy_Sioulas}}

\bibliography{sample631}
\bibliographystyle{aasjournal}

\end{document}